%% file: firstpassagetime1.tex
\documentclass[journal]{IEEEtran}

\usepackage{amssymb}
\usepackage[cmex10]{amsmath}
\usepackage{stmaryrd}
\usepackage{enumerate}
\usepackage{flushend}
\usepackage{tikz}
\usepackage{pgfplots}
\usepgflibrary{shapes}
\usetikzlibrary{arrows,shapes,chains,matrix,positioning,scopes}

\newtheorem{theorem}{Theorem}
\newtheorem{lemma}{Lemma}
\newtheorem{example}{Example}
\newtheorem{proposition}{Proposition}
\newtheorem{corollary}{Corollary}
\newtheorem{assumption}{Assumption}
\newtheorem{remark}{Remark}

\DeclareMathOperator*{\argmax}{\,arg\ max}

\newlength\tikzheight
\newlength\tikzwidth

\begin{document}

\title{First-Passage Time and Large-Deviation Analysis for Erasure Channels with Memory}

\author{\IEEEauthorblockN{Santhosh Kumar, Jean-Francois Chamberland, Henry D. Pfister}
\thanks{This material is based upon work supported by the National Science Foundation (NSF) under Grants No.~0747363 and No.~0830696.
Any opinions, findings, conclusions, and recommendations expressed in this material are those of the authors and do not necessarily reflect the views of the National Science Foundation.
This work was presented in part at Allerton Conferences on Communication, Control, and Computing, 2010 and 2011.

The authors are with the Department of Electrical and Computer Engineering, Texas A\&M University, College Station, TX 77843, USA (emails: santhosh.kumar@tamu.edu; chmbrlnd@tamu.edu; hpfister@tamu.edu).}
}

\maketitle

\begin{abstract}
This article considers the performance of digital communication systems transmitting messages over finite-state erasure channels with memory.
Information bits are protected from channel erasures using error-correcting codes; successful receptions of codewords are acknowledged at the source through instantaneous feedback.
The primary focus of this research is on delay-sensitive applications, codes with finite block lengths and, necessarily, non-vanishing probabilities of decoding failure.
The contribution of this article is twofold.
A methodology to compute the distribution of the time required to empty a buffer is introduced.
Based on this distribution, the mean hitting time to an empty queue and delay-violation probabilities for specific thresholds can be computed explicitly.
The proposed techniques apply to situations where the transmit buffer contains a predetermined number of information bits at the onset of the data transfer.
Furthermore, as additional performance criteria, large deviation principles are obtained for the empirical mean service time and the average packet-transmission time associated with the communication process.
This rigorous framework yields a pragmatic methodology to select code rate and block length for the communication unit as functions of the service requirements.
Examples motivated by practical systems are provided to further illustrate the applicability of these techniques.
\end{abstract}


\section{Introduction}
\label{section:Introduction}

Contemporary communication systems must be designed to accommodate the various applications that compose today's digital landscape.
In particular, mobile devices must meet the heterogeneous needs of various data flows in terms of delay tolerance and bandwidth requirements.
On the Internet backbone, congestion is often prevented by over-provisioning.
The large throughput and low latency of parallel optical lines provide a pragmatic solution that offers adequate network performance.
This approach, combined with localized content distribution networks and edge throttling, is key in supporting delay-sensitive traffic over the Internet core.
Unfortunately, a similar strategy cannot be applied to connect untethered devices, as wireless physical resources are limited and costly.
The narrow usable spectrum and the broadcast nature of wireless environments limit the effective bandwidth of wireless access networks and, hence, demand the efficient management of available resources.

In this article, we develop a mathematical framework that enables the optimal allocation of link resources for wireless systems in the context of delay-sensitive communication.
Distinguishing features of the proposed methodology include the joint treatment of finite-state channels with memory and queueing behavior at the transmitter.
The focus is on the first-passage time to an empty queue, and the methodology implicitly provides a distribution for the time it would take an additional packet to reach the head of the queue.
This view is not only important for resource allocation and performance evaluation, it offers a foundation for choosing among possible routes and distinct interfaces.
From an abstract perspective, we introduce a formulation where time-dependencies in channel states and decoding failures are captured meticulously.
In contrast to block-fading models, this formulation allows the seamless optimization of parameters such as code rate and block length.
This is instrumental in better understanding how these parameters affect the overall performance of delay-sensitive wireless connections.

Several contributions on the interplay between decisions at the physical layer and overall performance at the link layer can be found in the literature \cite{Gallager0471290483,Negi2002tit,Turin2002twcom,Bettesh2006tit}.
Notable approaches include the outage capacity \cite{Caire1999tit,KittEliaJavi2009TIT}, a probabilistic performance criterion based on the marginal distribution of channel blocks; the effective capacity \cite{Chang1994tac,Wu2003twcom} which captures the decay rate in buffer occupancy at the transmitter; and finite block-length analyses of wireless connections \cite{Haya2009TIT,polyanskiy2011dispersion}.
Physical resources can be optimized to reduce average delay by carefully selecting advantageous modulation schemes and coding strategies \cite{LiuZhouGian2005TWCOM,WangLiuGian2007TVT}.
Multi-objective problem formulations have also been explored.
For instance, the optimal tradeoff between power and delay has received attention in the past~\cite{Berry2002tit}.
The joint treatment of queueing and error-control coding has been examined by simultaneously considering the effective capacity of a link and the error exponent of a code family \cite{GoelNegi2004Globecom,GoelNegi2006ISIT}.
Markov models have been successfully employed in the queueing analysis of communication links with automatic repeat request \cite{Fant1996TVT,ElAzAltm2003PWC}.
Finally, powerful asymptotic techniques based on large deviations and heavy traffic limits have been developed to handle real-time traffic over unreliable links \cite{Kushner0387952640,Wu2006tac}.

This study differs from previous contributions in that it relates queueing behavior, error control coding and channel evolution without resorting to asymptotically long coding delays or rough approximations.
Decoding performance at the receiver captures channel correlation within a block, while the queueing aspect of the problem is key in understanding the impact of time-dependencies among successive decoding attempts.
Together, they provide an accurate assessment of overall system performance and lead to novel guidelines about efficient designs.

Furthermore, by focusing on the first-passage time to an empty queue~\cite{Norris0521633966}, we are able to bypass the search for representative arrival processes.
Rather, resource management can be performed adaptively based on current system conditions.
Having a distribution for the hitting time to an empty buffer enables the computation of several pertinent performance criteria such as the probability of violating a completion deadline, the mean first-passage time to an empty queue, and Chernoff bounds.
The proposed methodology is closely related to generating functions~\cite{graham1994concrete} and it works well for reasonably small initial buffer sizes, which are typical of communication systems subject to stringent delay restrictions.
On the other hand, under large buffers, this technique becomes somewhat cumbersome.
In this latter case, analyzing the large deviations governing the evolution of the system offers a promising new direction to derive meaningful guidelines for resource allocation and the selection of system parameters.
Indeed, the concentration of empirical measures can be used to gracefully adjust delay-sensitivity to the needs of real-time data flows by selecting the deviation threshold, i.e., the argument of the rate function~\cite{Dembo0387984062}.
Once a threshold is set, system parameters can be optimized according to this objective function and the resulting performance can be predicted accurately.

Throughout, we assume the availability of reliable acknowledgements using periodic feedback.
We also assume that the transmitter and receiver share a common randomness, which permits the utilization of random binary codes.
The remainder of this article is organized as follows.
Section~\ref{section:SystemModel} presents the channel model and the random coding scheme.
The queueing aspect of the problem is developed in Section~\ref{section:QueueingModel}.
A large deviations perspective on the mean transmission time and the average service rate is offered in Section~\ref{section:LDP}.
The findings are supplemented by a discussion of pertinent criteria for performance evaluation, along with numerical examples.
Concluding remarks and possible avenues of future research are exposed in Section~\ref{section:Conclusions}.

\section{System Model}
\label{section:SystemModel}

One physical aspect of wireless communication that we are particularly interested in is channel memory.
From a queueing perspective, it is well known that correlation over time can drastically alter the stationary distribution of a queueing system \cite{Kleinrock0471491101,Gross047179127X}.
In a similar manner, channel memory can have a strong impact on overall performance, as it induces time-dependencies in the service process at the transmitter.
This phenomenon is especially important for delay-sensitive applications that require the reliable, ordered delivery of data streams.
A prime model class in dealing with such dependencies is composed of finite-state channels with memory \cite{gilbert1960capacity,elliott1963estimates,Cover0471062596}.
System models derived from this class of channels are typically mathematically tractable, and they offer a natural mechanism to account for correlation over time.
Moreover, insights acquired by studying erasure channels can often be translated to error channels or, at least, provide partial intuition about promising solutions for the latter, more challenging scenarios.

This article revolves around a communication paradigm where information bits flow from a source to a destination.
The transmitter is assumed to possess a message of a certain length at the onset of the data transfer, and forward error correction is employed to shield content from potential symbol erasures.
At the beginning of a transmission, the leading information bits stored at the source are grouped into a segment, and redundancy is added to this message using block encoding.
The resulting codeword is then sent over a finite-state erasure channel with memory.
Contingent upon the channel realization, the destination can either retrieve the data contained in the transmitted codeword or it declares a decoding failure.
Successful transmissions are acknowledged and the corresponding bits are then discarded from the source buffer.
Otherwise, the leading information bits remain in the queue.
We emphasize that, in this framework, the original data sequence is guaranteed to be transferred unaltered.
However, the completion time of the queue-emptying process is a random variable that depends on the coding/decoding strategy adopted and on the realization of the channel.

\subsection{Channel Abstraction}

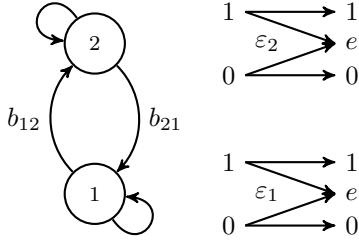
\begin{figure}[bth]
\begin{center}
\input{Figures/erasurechannel}
\caption{Communication at the bit level takes place over a finite-state erasure channel with memory.
While in state $i$, the probability of a bit erasure is $\varepsilon_i$.
The evolution of the channel over time forms a Markov process.}
\label{figure:FiniteStateChannel}
\end{center}
\end{figure}

As indicated above, we capture channel stochasticity and its impact on the communication link using a finite-state Markov process.
Several pertinent communication scenarios can be modeled in this manner \cite{Wang1995tvt,zhang1999finite,sadeghi2008finite}.
At a particular time instant, we assume that the channel can be in one of $k$ states taking value in $\mathcal{C}=\{1, 2, \ldots, k\}$.
State transitions over time form a Markov chain.
We denote the corresponding transition probability matrix by
\begin{equation*} 
\mathbf{B} =
\begin{bmatrix}
 b_{11} & b_{12} & \cdots & b_{1k} \\
 b_{21} & b_{22} & \cdots & b_{2k} \\
 \vdots & \vdots &  \ddots & \vdots \\
 b_{k1} & b_{k2} & \cdots & b_{kk}
\end{bmatrix} .
\end{equation*}
Entry $b_{ij}$ in matrix $\mathbf{B}$ represents the conditional probability that, starting from state~$i$, the channel transitions to state~$j$.
As such, $\mathbf{B}$ is a right stochastic matrix.
When in state $i$, the transmitted symbol is erased with probability $\varepsilon_i$ and, consequently, it is received correctly with probability $1-\varepsilon_i$.
For notational convenience, we impose a quality ordering on the channel states, i.e., $\varepsilon_i \geq \varepsilon_{j}$ whenever $i < j$.
We represent the state of the channel at time instant $n$ by $C_n$.
We note that $\{ C_n \}$ is a first-order Markov process.
A diagram illustrating the operation of the communication link for a two-state channel appears in Fig.~\ref{figure:FiniteStateChannel}.

\begin{assumption} \label{assumption:PrimitiveMarkovChannel}
Throughout, we hypothesize that the chain governing the finite-state channel is irreducible and aperiodic.
We also assume that this Markov channel is non-trivial in that there exists a state $i \in \mathcal{C}$ such that $\varepsilon_i < 1$.
\end{assumption}

As we shall see, these conditions guarantee the existence of a random coding scheme for which the transmission process terminates in finite time, almost surely.
These transmission schemes are the only ones of interest for our purpose.
In that sense, Assumption~\ref{assumption:PrimitiveMarkovChannel} is introduced to prevent difficulties that arise from idiosyncratic, irrelevant scenarios.

\subsection{Coding Scheme}

The envisioned system employs forward error correction to counteract possible channel erasures.
A codeword transmission attempt is initiated by selecting the leading $K$ bits from the source buffer.
Redundancy is then added to this data segment through the encoding process.
A random coding scheme is adopted as a mathematically convenient abstraction to realistic implementations \cite{Gallager0471290483,Richardson0521852293}.
To create each codeword transmission, a random binary parity check matrix of size $(N-K) \times N$ is generated.
Every entry is selected uniformly over the binary alphabet, independently from other elements.
The resulting codebook corresponds to the nullspace of this matrix.
Such a coding scheme ensures that successful decoding of different codewords are conditionally independent given the channel states at the respective transmission times.
This will greatly simplify the ensuing analysis.
We assume that maximum-likelihood decoding is performed at the receiver.

We emphasize that this mode of operation requires shared randomness at the source and the destination.
Interestingly, this coding scheme is known to perform well for large block lengths; and it supports flexible rates of communication, any rate of the form $K/N$ where $0 \leq K \leq N$ is admissible.
These random codes have the additional property that the average probability of decoding failure depends only on the number of erasures caused by the channel and not on the specific locations of these erasures.
Provided that $e$ erasures have occurred during transmission, the probability of decoding failure can be evaluated explicitly,
\begin{equation}
\label{equation:DecodingFailure}
 P_{\mathrm{f}} (N-K,e)=1- \prod_{l=0}^{e-1} \left( 1-2^{l-(N-K)} \right).
\end{equation}
A proof for this statement is based on the equivalence between the linear independence of the $e$ erased columns in the parity check matrix and the event of a successful decoding~\cite{Richardson0521852293}.
Throughout this article, $P_{\mathrm{f}}(p,e)$ denotes 
\begin{align}
  \label{equation:DecodingFailureExtended}
  P_{\mathrm{f}} (p , e)=
  \begin{cases}
    1 - \prod_{l=0}^{e - 1} \left( 1 - 2^{l - p} \right) & \text{if $e \leq p$} \\
    1 & \text{if $p < e \leq N$}
  \end{cases}
\end{align}
which is the average probability of decoding failure under maximum likelihood of a codebook generated by using a random binary parity check matrix of size $p \times N$, for any $N \geq p$, when $e$ erasures have occurred.

\subsection{Distribution of Erasures}

From the discussion above, we gather that the number of erasures suffered by a codeword plays a critical role in determining overall system performance, as it dictates the probability of decoding failure.
This random variable thus warrants due attention.
Let $E$ denote the number of erasures occurring in a given packet transmission.
Since the probability of decoding failure of a codeword depends only on the number of erasures, it suffices to consider probabilities of the form $\Pr(E=e,C_{N+1}=j | C_1=i)$ to characterize the evolution of the system.
Note that $C_1$ and $C_{N+1}$ correspond to the channel state transitions across the first codeword transmission.
We can describe this distribution in a compact form using matrix generating functions.
Define matrix $\mathbf{B}_x$ by
\begin{equation*}
\mathbf{B}_x=
 \begin{bmatrix}
  b_{11}(1-\varepsilon_{1}+\varepsilon_{1}x) & \cdots & b_{1k}(1-\varepsilon_{1}+\varepsilon_{1}x) \\
  b_{21}(1-\varepsilon_{2}+\varepsilon_{2}x) & \cdots & b_{2k}(1-\varepsilon_{2}+\varepsilon_{2}x) \\
  \vdots & \ddots & \vdots \\
  b_{k1}(1-\varepsilon_{k}+\varepsilon_{k}x) & \cdots & b_{kk}(1-\varepsilon_{k}+\varepsilon_{k}x) 
\end{bmatrix}.
\end{equation*}
Throughout this article, $\llbracket x^n \rrbracket$ denotes the linear operator that maps a polynomial in $\Re [x]$ to the coefficient of $x^n$.
For $e \in \mathbb{N}_0$ and $i,j \in \mathcal{C}$, one can show that~\cite{graham1994concrete}
\begin{equation}
\label{equation:NumberOfErasures}
 \Pr(E=e,C_{N+1}=j | C_{1}=i) = \llbracket x^e \rrbracket \left[ \mathbf{B}_x^{N} \right]_{i,j}
\end{equation}
where, in this case, $E$ denotes the number of erasures over an interval of length~$N$.
The probability that Markov process $\{C_n\}$ coincides with a specific sequence of states is equal to the probability of a certain path through the matching trellis.
Moreover, at each point in time, the probability of observing an erasure only depends on the current state.
Consequently, taking the $N$th power of matrix $\mathbf{B}_x$ is an efficient way to compute the aggregate conditional probability of observing exactly $e$ erasures, given an initial probability distribution and an end state.
In other words, $\mathbf{B}_x^{N}$ offers a way to simultaneously sum all the relevant paths through the trellis.
It is also possible to compute such probabilities through nested sums~\cite{Wilhelmsson-com99}, but the ensuing equations rapidly become cumbersome for large values of $N$ and Markov chains with sizable state spaces.
 
Given initial state~$i$ and for a fixed final state~$j$, we can apply the total probability theorem to compute the probability of decoding failure,
\begin{equation}
\label{equation:ConditionalProbabilityDecodingFailure}
\sum_{e=0}^N P_{\mathrm{f}}(N-K,e) 
\Pr \left( E=e, C_{N+1} = j | C_1 = i \right) .
\end{equation}
These conditional probabilities, along with the progression of the channel states, underlie the evolution of the queueing system.

\begin{remark} \label{remark:NonTrivialCodes}
As a side note, it is instructive to point out that, under Assumption~\ref{assumption:PrimitiveMarkovChannel}, there exist values for $N$ and $K$ such that the probability of decoding success as a function of $C_1$ is not uniformly zero.
In particular, if $i$ is a channel state such that $\varepsilon_i < 1$, then for large enough $N$ and $N-K$, the probability of decoding failure in \eqref{equation:ConditionalProbabilityDecodingFailure} will be less than one.
Random codes for which the conditional probability of decoding success is not uniformly zero are termed non-trivial.
\end{remark}

\section{Queueing Model}
\label{section:QueueingModel}

This section describes the queueing behavior of our system.
First, we assume that the number of information bits present at the source at the beginning of the communication process is fixed and equal to $\ell$.
Given a code rate and block length, the source takes the leading $K$ data bits and encodes the resulting segment into a codeword of length $N$ using the scheme described in the preceding section.
This codeword is then sent to the destination through $N$ consecutive uses of the erasure channel.
A service opportunity occurs every time the random code and channel realization jointly permit reliable decoding.
We emphasize, again, that the destination is assumed to possess the ability to acknowledge the successful reception of codewords through instantaneous feedback.
As such, the selected information bits remain in the transmit queue until a corresponding codeword is decoded faithfully at the destination.
This data segment is immediately discarded from the buffer upon successful decoding of a packet.

In its simplest form, this scheme represents a variation of automatic repeat request (ARQ).
We note that this mode of operation is somewhat na\"{i}ve in that the information contained in failed decoding attempts is disregarded.
A more astute implementation will seek to leverage past failures by performing joint decoding over all the observed messages pertaining to the current data segment.
Incremental redundancy and hybrid automatic repeat request are valuable techniques that can improve performance \cite{Comroe1984tcom,sesia2004incremental,le2007queueing}.
In this article, we discuss both ARQ and its hybrid variant, where partial information from failed transmission attempts is incorporated in the decoding process.
Still, we focus largely on the rudimentary scheme because it admits a simpler, more elegant characterization while preserving the natural tradeoff between error protection and payload content.
Overall, the proposed methodology yields pertinent results that help improve our understanding of delay-sensitive systems.

Our primary interest lies in the distribution of the time elapsed until the message originally contained in the source buffer becomes wholly available at the destination.
To capture this quantity adequately, we need to examine the evolution of the queue.
The length of the queue can be expressed in terms of the number of data segments awaiting transmission.
If a queue initially contains $\ell$ information bits, then it will require the successful reception of $m=\lceil  \ell/K \rceil$ codewords until the last segment gets processed.
The number of segments in the transmit buffer therefore becomes a measure of residual work until our objective is met, and it is intrinsically linked to the state of our communication system.

Codeword $s$ denotes the block of transmitted bits during the time instants $sN+1,\ldots,(s+1)N$, where $s \geq 0$.
These codewords include both decoding successes and failures.
For $N$ fixed, we denote the size of the queue at the onset of codeword $s$ by $Q_s$.
We note that the state of the bit-erasure channel at the same time instant is $C_{sN+1}$.
Thus at the onset of the first codeword transmission ($s=0$), the size of the queue is $Q_0$ and the state of the bit-erasure channel is $C_1$.
The rapid succession of symbols in the bit-erasure channel compared to events taking place in the queue produces the mismatch in indexing between $Q_s$ and $C_{sN+1}$.
Indeed, queue transitions are only possible at the completions of decoding attempts, which only occur after every $N$ symbol transmissions.
The resulting stochastic process $\{ Q_s \}$ is a hidden Markov process, as it is determined partly by the evolution of the unobserved channel process $\{ C_n \}$.
While $\{ Q_s\}$ alone does not possess the Markov property, it is possible to create an augmented process containing $Q_s$ with this desirable attribute.
The particulars of the procedure depend on whether one is considering the standard ARQ framework or its hybrid variant.
We treat these two instances separately.

\subsection{Automatic Repeat Request}
\label{section:AutomaticRepeatRequest}

As the title suggests, this section focuses exclusively on the scenario where the source and the destination employ ARQ to overcome channel erasures and, thereby, achieve reliable data transmission.
In particular, the information contained in past decoding attempts is disregarded by the decoder when receiving the latest codeword.
To build a suitable model, we consider the random vector $U_s = \left( C_{sN+1}, Q_s \right)$ composed of channel state and queue length.
We wish to show that this vector contains all the relevant information to track the evolution of the system.

\begin{theorem} \label{theorem:MarkovProperty1}
The aggregate process $\{U_s\}_{s \geq 0}$ possesses the Markov property.
That is, conditioned on $U_t = (i, q)$, the stochastic process $\{ U_{s + t} \}_{s \geq 0}$ is independent of $U_0, \ldots, U_{t-1}$.
\end{theorem}
\begin{IEEEproof}
See Appendix \ref{appendix:MarkovProperty1}.
\end{IEEEproof}

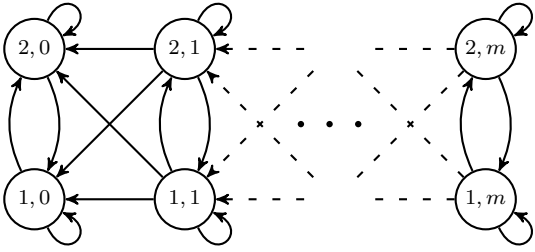
\begin{figure}[bth]
\begin{center}
\input{Figures/packeterasurequeue}
\caption{This figure illustrates the progression of the queueing system for a service process that is governed by a two-state Markov erasure channel.
System states, which are composed of queue lengths and channel states, are represented by circles.
Admissible transitions are marked by the arrows.}
\label{figure:PacketErasureQueue}
\end{center}
\end{figure}

Using the total probability theorem, we can write the transition probabilities of $\{ U_s \}$ as follows,
\begin{equation} \label{equation:StateTranstionProbabilities}
\begin{split}
\Pr & \left( U_{s+1} = (j, q_{s+1}) | U_s = (i, q_s) \right) \\
&= \sum_{ e = 0 }^{ N }
\Pr \left( Q_{s+1} = q_{s+1} | E=e, Q_s = q_s \right) \times \\
&\quad \Pr \left( E=e, C_{(s+1)N+1} = j | C_{sN+1} = i \right)
\end{split}
\end{equation}
where $i, j \in \mathcal{C}$.
For a non-empty queue, the first part of each summand corresponds to one of three possible cases,
\begin{equation*}
\begin{split}
\Pr & \left( Q_{s+1} = q_{s+1} | E=e, Q_s = q_s \right) \\
&= \begin{cases}
P_{\mathrm{f}} (N - K,e), & q_{s+1} = q_s \\
1 - P_{\mathrm{f}}(N - K,e), & q_{s+1} = q_s - 1 \\
0, & \text{otherwise} .
\end{cases}
\end{split}
\end{equation*}
The probability of decoding failure $P_{\mathrm{f}} (\cdot, \cdot)$ appears in \eqref{equation:DecodingFailure}, while the conditional distribution of erasures within a block is given in \eqref{equation:NumberOfErasures}.
Thus, we have already developed the tools necessary to efficiently compute the value of every transition probability in~\eqref{equation:StateTranstionProbabilities}.
The evolution of the queueing system and its admissible transitions are depicted graphically in Fig.~\ref{figure:PacketErasureQueue}.
The states $\{ (\cdot, q) \}$ are collectively referred to as the $q$th level of the queue.
The first-passage time to an empty buffer is therefore equivalent to the hitting time to level zero.
Due to the repetitive structure of this augmented system, the hitting time to a lower level will play a key role in finding a tractable solution to the problem at hand.

An additional quantity of interest in the analysis of delay-sensitive systems is the mean service rate.
To compute this quantity, it is convenient to analyze the service process $\{ D_s \}$, where $D_s$ indicates the potential of a successful decoding event at time~$s$, $s \geq 0$.
That is, $D_s = 1$ when a message can (or could) be decoded faithfully at the destination; and $D_s = 0$ otherwise.
In words, the sequence $\{ D_s \}$ indicates time instants at which blocks of information can be transferred successfully to the destination.
As in the case of the queueing abstraction, the stochastic process $\{ D_s \}$ forms a hidden Markov process which can be lifted to an augmented Markov process.
Let $V_s = \left( C_{(s+1)N+1}, D_s \right)$ denote a random vector composed of the state of the erasure channel at the onset of block $s+1$, together with the indicator of a service opportunity during block~$s$.
As in Theorem~\ref{theorem:MarkovProperty1}, one can show that the stochastic process $\{ V_s \}$ forms a Markov chain.

We note that the transition probabilities of $\{ D_s \}$ are closely related to those of $\{Q_s\}$.
Since there are no arrivals in our framework, the evolution of these processes are governed by
\begin{equation*}
Q_{s+1} = \left( Q_s - D_s \right)^+ .
\end{equation*}
For convenience, we establish a succinct notation for the transition probabilities of our two augmented processes,
\begin{equation} \label{equation:AugmentedTransitionProbabilities}
\begin{split}
\kappa_{ij} &= \Pr ( U_{s+1} = (j, q) | U_s = (i, q) ) \\
&= \Pr ( V_{s+1} = (j, 0) | V_s = (i, d) ) \\
\mu_{ij} &= \Pr ( U_{s+1} = (j, q-1) | U_s = (i, q) ) \\
&= \Pr ( V_{s+1} = (j, 1) | V_s = (i, d) )
\end{split}
\end{equation}
where  $q \in \mathbb{N}$, $i, j \in \mathcal{C}$ and $d \in \{ 0, 1\}$.
These common definitions draw further attention to the close connection between $\{ U_s \}$ and $\{ V_s \}$.

In view of Remark~\ref{remark:NonTrivialCodes} and for non-trivial codes, there exists $i \in \mathcal{C}$ such that $\mu_{ij} > 0$.
This implies that the states associated with an empty buffer form the only closed communicating class and, as such, the remaining states are transient~\cite{Norris0521633966}.
Since the number of states in the augmented chain is finite, this structure ensures that the task of emptying the transmit buffer is carried out in finite time, almost surely.

The symmetric decomposition of the queueing system into levels suggests an approach  based on the quasi-birth-death structure of the chain.
Suppose that the buffer contains exactly $m$ data segments at time zero, i.e., $Q_0 = m$.
We can define the hitting time from level~$m$ to level~$q$ of the chain as
\begin{equation} \label{equation:HittingTimes}
H_q = \inf \{ s \geq 0 | Q_s = q \} ,
\end{equation}
where $0 \leq q < m$.
That is, $H_q$ designates the time instant at which the process $\{ U_s \}$ first enters the $q$th level of the queue.
We emphasize that, under the mild assumptions discussed above, $H_q$ is almost surely finite.
For consistency, we also define $H_m = 0$.
Noting that $Q_s$ is a non-increasing process, we can write the sojourn time at level~$q$ as
\begin{equation*}
T_q = H_{q-1} - H_q ,
\end{equation*}
where $0 < q \leq m$.
That is, random variable $T_q$ denotes the amount of time $\{ U_s \}$ stays at level~$q$ before leaving for the subsequent lower level.

We are especially interested in $H_0$, the first-passage time to an empty queue.
Taking advantage of the structure of the augmented Markov chain, we can fragment $H_0$ into a sum of elementary components.
Specifically, the hitting time $H_0$ is equal to the sum of the sojourn times $T_1, \ldots, T_m$, i.e.,
\begin{equation*}
H_0 = \sum_{q=1}^m T_q .
\end{equation*}
The sojourn times $T_q$ and $T_{q-1}$ are coupled through the channel state $C_{N H_{q-1} + 1}$ and hence are not independent.
However, since the codebooks over different codeword transmissions are independent, the sojourn times $T_1,\ldots,T_m$ are conditionally independent given the channel states $\left\{ C_{ N H_q + 1} \right\}_{q=0}^m$.
The sojourn times $T_1,\ldots,T_m$ are also conditionally identically distributed.
That is,
\begin{align*}
  \Pr\left(T_q=t,C_{N H_{q-1} + 1}=j | C_{N H_{q} + 1}=i \right)
\end{align*}
is independent of $q$.
A powerful means to compute the distribution of $H_0$ is to employ generating functions extended to matrices \cite{graham1994concrete}, exploiting the conditional independence and the identical distribution among the sojourn times $\{ T_q \}$.
This more intricate version of the generating function is necessary to keep track of the channel state entered after each downward queue transition.
This method is described below.

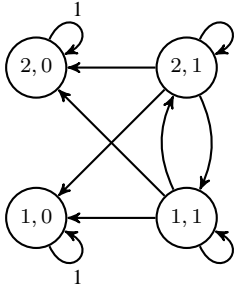
\begin{figure}[tb]
\begin{center}
\input{Figures/packeterasureblock}
\caption{This reduced Markov diagram represents one of the quasi-birth-death subcomponents of the queueing system.
Starting from any distribution over these four states, it is possible to characterize the sojourn time $T$ spent at level one.
This is a key step in deriving the first-passage time to an empty buffer.}
\label{figure:PacketErasureBlock}
\end{center}
\end{figure}

Consider a reduced Markov chain composed of states $\{ (i, 0), (i, 1) \}_{i=1}^{k}$, as shown in Fig.~\ref{figure:PacketErasureBlock} for a Gilbert-Elliott channel.
This reduced Markov chain represents one downward queue transition of the original system.
Under proper state ordering, we can write the transition probability matrix for the reduced subsystem as
\begin{equation} \label{equation:ReducedSubsystemMatrix}
\mathbf{P} = \begin{bmatrix} \mathbf{I} & \mathbf{0} \\
\mathbf{M} & \mathbf{K} \end{bmatrix} ,
\end{equation}
where we have implicitly defined matrices
\begin{xalignat*}{2}
\mathbf{M} &= \begin{bmatrix}
\mu_{11} & \cdots & \mu_{1k} \\
\mu_{21} & \cdots & \mu_{2k} \\
\vdots & \ddots & \vdots \\
\mu_{k1} & \cdots & \mu_{kk}
\end{bmatrix} &
\mathbf{K} &= \begin{bmatrix}
\kappa_{11} & \cdots & \kappa_{1k} \\
\kappa_{21} & \cdots & \kappa_{2k} \\
\vdots & \ddots & \vdots \\
\kappa_{k1} & \cdots & \kappa_{kk}
\end{bmatrix} .
\end{xalignat*}
We emphasize that $\mathbf{P}$ is a stochastic matrix.
As a consequence of the Perron-Frobenius theorem, we know that the spectral radius associated with $\mathbf{P}$ is one~\cite{Horn0521386322}.

Define sojourn time $T$ as the time spent at queue-level~$1$ of the reduced Markov chain.
Mimicking our original notation, let $Q_s$ denote the level of the queue (either $1$ or $0$) at the onset of codeword $s$ and let $U_s=(C_{sN+1},Q_s)$.
Suppose the reduced Markov chain starts at queue-level $1$, i.e.~$Q_0=1$, then
\begin{align*}
  T=\inf \left\{ s \geq 0 | Q_s = 0 \right\} .
\end{align*}
The random variables $\{T_q\}_{q=1}^{m}$ and $T$ have identical conditional distributions.
That is, for any $1 \leq q \leq m$,
\begin{align*}
  & \Pr \left( T=t, C_{NT+1}=j | C_{1}=i  \right) \\
  & \quad = \Pr \left(T_q=t,C_{N H_{q-1} + 1}=j | C_{N H_{q} + 1}=i \right) .
\end{align*}

The distributions of the sojourn times $T_1,\ldots,T_m$ are important for determining the distribution of $H_0$.
Thus, the above relation between $T_1,\ldots,T_m$ and $T$ implies that the distribution of $T$ is critical.
Generating functions are an elegant way to characterize such distributions.
Define matrix generating function $\mathbf{G}_{T}(z)$ entrywise by
\begin{equation} \label{equation:GeneratingMatrix}
\left[ \mathbf{G}_{T}(z) \right]_{ij}
= \mathrm{E} \left[ z^{T} \mathbf{1}_{ \{ C_{NT+1} = j \} } | C_1 = i \right]
\end{equation}
where $\mathbf{1}_{\{ \cdot \}}$ is the standard set indicator function.

\begin{lemma} \label{lemma:BitErasureGeneratingMatrix}
For the reduced subsystem associated with \eqref{equation:ReducedSubsystemMatrix}, the matrix generating function $\mathbf{G}_T (z)$ is equal to
\begin{equation} \label{equation:GeneratingMatrixSpecific}
\mathbf{G}_T (z)
= \left( \mathbf{I} - \mathbf{K} z \right)^{-1} \mathbf{M} z .
\end{equation}
\end{lemma}
\begin{IEEEproof}
The matrix generating function $\mathbf{G}_T (z)$ can be obtained by treating the entries of $\mathbf{P}$ as real polynomials in $z$, with
\begin{equation*}
\mathbf{P}_z = \begin{bmatrix} \mathbf{I} & \mathbf{0} \\
\mathbf{M} z & \mathbf{K} z \end{bmatrix} .
\end{equation*}
Consider the two states $(i, 1)$ and $(j, l)$, where $l=0$ or $l=1$.
Their indices in the ordering associated with $\mathbf{P}$ are $k+i$ and $lk+j$, respectively.
Recall that $\llbracket z^t \rrbracket$ denotes the operator that maps a polynomial in $z$ to the coefficient of $z^t$.
Suppose that, at time zero, the reduced system starts in state $(i,1)$.
After $s$ transmissions, the reduced system will be in state $(j,1)$ only when all the $s$ transmissions result in decoding failures.
Thus
\begin{align}
  \label{equation:reduced_system_transition_1}
  \Pr \left( U_s=(j,1) | U_0=(i,1) \right) = \left[ \mathbf{K}^{s} \right]_{i,j} .
\end{align} 
Similarly, the probability that the reduced system is in state $(j,0)$ after $s$ transmissions and having spent exactly $t$ steps in queue-level~$1$, where $1 \leq t \leq s$, is given by
\begin{align*}
  \sum_{h=1}^{k} \Pr \left( U_s=(j,0), U_{t}=(j,0), U_{t-1}=(h,1) | U_0=(i,1) \right) .
\end{align*}  
Since the reduced system does not transition to a different state after reaching queue-level~$0$ (see Fig.~\ref{figure:PacketErasureBlock}), this can be reduced to 
\begin{align}
  & \sum_{h=1}^{k} \Pr \left( U_{s}=(j,0), U_{t}=(j,0), U_{t-1}=(h,1) | U_0=(i,1) \right) \notag \\
  & \qquad = \sum_{h=1}^{k} \Pr \left( U_{t}=(j,0), U_{t-1}=(h,1) | U_0=(i,1) \right) \notag \\
  & \qquad = \left[ \mathbf{K}^{t-1} \mathbf{M} \right]_{i,j} . \label{equation:reduced_system_transition_2}
\end{align}
Combining (\ref{equation:reduced_system_transition_1}) and (\ref{equation:reduced_system_transition_2}), the joint probability that the reduced system is in state $(j, l)$ at time $s > 0$ and has spent exactly $t$ steps at queue-level~$1$, where $1 \leq t \leq s$, can be expressed compactly as
\begin{equation*}
\Pr (S_s = t, U_s = (j, l) | U_0 = (i, 1))
= \llbracket z^t \rrbracket \left[ \mathbf{P}_z^s \right]_{k+i, lk+j} ,
\end{equation*}
where $S_s$ represents the total time spent at queue-level~$1$ over the interval from zero to instant~$s$.
Since $T$ is a discrete random variable that is finite almost surely,
\begin{align*}
  \allowdisplaybreaks
  \left[ \mathbf{G}_{T}(z) \right]_{ij} &= \mathrm{E} \left[ z^{T} \mathbf{1}_{ \{ C_{NT+1} = j \} } | C_1 = i \right] \\
  &= \lim_{s \rightarrow \infty} \sum_{t=0}^{s} \Pr \left( T=t, C_{Nt+1}=j | C_1=i \right) z^t \\
  &= \lim_{s \rightarrow \infty} \sum_{t=0}^{s} \Pr \left( S_s=t, U_s\!=\!(j,0) | U_0\!=\!(i,1) \right) z^t \\
  &= \lim_{s \rightarrow \infty} \sum_{t=0}^{s} \left( \llbracket z^t \rrbracket \left[ \mathbf{P}_z^s \right]_{k+i,j} \right) z^t \\
  &=\lim_{s \rightarrow \infty} \left[ \mathbf{P}_z^s \right]_{k+i,j} .
\end{align*}
Therefore the generating matrix $\mathbf{G}_T (z)$ can be obtained as
\begin{align*}
  \mathbf{G}_T (z) &= \lim_{s \rightarrow \infty}
  \begin{bmatrix} \mathbf{0} & \mathbf{I} \end{bmatrix} 
  \mathbf{P}_z^s  
  \begin{bmatrix} \mathbf{I} \\ \mathbf{0} \end{bmatrix} \\
  & = \lim_{s \rightarrow \infty}\begin{bmatrix} \mathbf{0} & \mathbf{I} \end{bmatrix} 
  \begin{bmatrix} \mathbf{I} & \mathbf{0} \\ \sum_{t=1}^{s} \mathbf{K}^{t-1}\mathbf{M} z^{t}  & \mathbf{M}^{s} z^{s}   \end{bmatrix}
  \begin{bmatrix} \mathbf{I} \\ \mathbf{0} \end{bmatrix} \\
  & = \lim_{s \rightarrow \infty} \sum_{t=1}^{s} \mathbf{K}^{t-1}\mathbf{M} z^{t} \\
  &= \left( \mathbf{I} - \mathbf{K} z \right)^{-1} \mathbf{M} z .
\end{align*}
The above equation holds for all $|z| < \varrho(\mathbf{K})^{-1}$, where $\varrho(\cdot)$ denotes the spectral radius of its matrix argument.
\end{IEEEproof}

\subsection{Hybrid Automatic Repeat Request}
\label{section:HybridAutomaticRepeatRequest}

Hybrid ARQ is a mechanism that seeks to incorporate the partial information contained in failed transmissions into the subsequent decoding attempts of the same data segment.
In this sense, it differs significantly from ARQ only when the initial decoding of a data segment fails.
For finite-state erasure channels with memory, the evolution of a hybrid ARQ system can be characterized completely, although in a somewhat cumbersome manner.
To implement hybrid ARQ with random codes, we must modify our coding strategy slightly.

Herein, we focus on hybrid schemes with finite depths.
That is, the transmitter-receiver pair has a predetermined number of tries to successfully transmit a data segment.
Our favored implementation relies on puncturing random codes.
In a way analogous to our previous approach, we generate a codebook by creating a random binary parity check matrix of size $(aN - K) \times aN$, where $a$ is the depth of the hybrid ARQ scheme.
Again, the entries are selected uniformly from the binary alphabet and the codebook is equal to the nullspace of this matrix.
The hybrid ARQ scheme progresses as follows.
First, an information segment is mapped to a codeword of length $aN$.
During the initial transmission, the leading $N$ symbols of this codeword are sent over the erasure channel.
Upon completion of this phase, the destination tries to recover the original data segment.
When decoding fails, the next $N$ symbols are sent and the aggregate message is run through a maximum-likelihood decoder.
This process continues, communicating $N$ symbols at a time, until the message is successfully decoded at the destination or the total number of attempts reaches its limit.

Since untransmitted symbols can be classified as erasures for the purpose of decoding, we can leverage \eqref{equation:DecodingFailureExtended} in assessing the probabilities of decoding failure at the destination.
That is, when $s$ codeword chunks are present at the destination, out of which a total of $e$ symbols are erased, the probability of decoding failure can be written as
\begin{equation}
\label{equation:DecodingFailureHARQ}
\begin{split}
P_{\mathrm{f}} & (aN-K, e+(a-s)N) \\
&= 1 - \prod_{i=0}^{e + (a-s)N -1} \left( 1-2^{i-(aN-K)} \right).
\end{split}
\end{equation}
Comparing this expression for $s=1$ and $a>1$ to \eqref{equation:DecodingFailure}, we gather that the probability of decoding failure after receiving one chunk of length~$N$ for the hybrid ARQ scheme differs from the probability of failure in standard ARQ.
Indeed, there is a slight penalty for the initial transmission resulting from using a random code tailored to hybrid ARQ.
The following proposition establishes a uniform bound on the loss in performance associated with the hybrid scheme.

\begin{proposition}
\label{proposition:PropertiesOfPf}
Suppose that $p$ and $e$ are fixed, positive integers.
The function of $n$ defined by
\begin{equation*}
P_{\mathrm{f}} (p + n, e + n)=
\begin{cases}
1 - \prod_{l=0}^{n + e - 1} \left( 1 - 2^{l - p - n} \right) & \text{if $e \leq p$} \\
1 & \text{if $e > p$}
\end{cases}
\end{equation*}
is monotone increasing.
Furthermore, the difference between this function and $P_{\mathrm{f}} (p, e)$ is uniformly bounded,
\begin{equation*}
P_{\mathrm{f}} (p + n, e + n) - P_{\mathrm{f}} (p, e)
\leq 2^{-p} .
\end{equation*}
\end{proposition}
\begin{IEEEproof}
See Appendix \ref{appendix:PropertiesOfPf}.
\end{IEEEproof}

The probability of decoding failure for the initial transmission of the hybrid ARQ scheme is $P_{\mathrm{f}}(aN-K,e+(a-1)N)$, and it is $P_{\mathrm{f}}(N-K,e)$ for the standard ARQ scheme when the codeword suffers $e$ erasures.
As an immediate consequence of Proposition~\ref{proposition:PropertiesOfPf}, we know that the penalty incurred in using hybrid ARQ in terms of probability of decoding failure at the first attempt is
\begin{align*}
  P_{\mathrm{f}}(aN-K,e+(a-1)N) - P_{\mathrm{f}}(N-K,e) \leq 2^{-(N-K)} ,
\end{align*}
which remains very small for typical scenarios.
This brings credibility to employing a punctured random code in our analysis.

Using random codes over erasure channels leads to some highly desirable properties for the hybrid ARQ problem.
These properties are, in turn, instrumental in finding expressions for the probabilities of success at intermediate decoding attempts.
Suppose that a codebook is generated using a $(aN - K) \times aN$ parity check matrix.
For this specific code, if decoding fails given the first $sN$ received symbols (including erasures), then it will necessarily be impossible to decode the message using the leading $(s-1)N$ received symbols.
This nesting is in stark contrast to error channels.

We employ $P_{\mathrm{f}}^{(s)}(j | i)$ and $P_{\mathrm{s}}^{(s)}(j | i)$ to denote the conditional probability of decoding failure and first reliable decoding success at attempt $s$, respectively, with final state~$j$ and given initial state~$i$.
The conditional probabilities of decoding failure are equal to
\begin{equation*}
\begin{split}
P_{\mathrm{f}}^{(s)} (j | i)
= \sum_{e=0}^{sN} & P_{\mathrm{f}}(aN-K,e+(a-s)N) \times \\
& \Pr \left( E_{sN}=e, C_{sN+1} = j | C_1 = i \right) .
\end{split}
\end{equation*}
Above, $E_{sN}$ represents the number of erasures over the discrete interval $[1, sN]$.
Given the probabilities of failure events, the conditional probabilities of success can be evaluated in a recursive fashion.
Since decoding failure and decoding success at attempt one are complementary events, we have
\begin{align*}
  \Pr(C_{N+1} = j | C_1 = i)=  P_{\mathrm{f}}^{(1)}(j | i) + P_{\mathrm{s}}^{(1)}(j | i) .
\end{align*} 
Thus, the probability of a success at time one with final state~$j$ given initial state~$i$, can be written as
\begin{equation*}
\begin{split}
&P_{\mathrm{s}}^{(1)} (j | i)
= \Pr (C_{N+1} = j | C_1 = i) - P_{\mathrm{f}}^{(1)} (j | i) .
\end{split}
\end{equation*}
We note that this equation is the complement of \eqref{equation:ConditionalProbabilityDecodingFailure}, with a convenient new notation and appropriate parameters.

Similarly, consider the first two attempts in a hybrid ARQ scheme.
Three disjoint events can occur: decoding failure at attempt two, decoding success for the first time at attempt two, decoding success at attempt one after which the channel enters some state $l$.
Summing over all intermediate states $l$,
\begin{align*}
& \Pr (C_{2N+1} = j | C_1 = i) = P_{\mathrm{f}}^{(2)} (j | i) + P_{\mathrm{s}}^{(2)} (j | i) \\ 
& \qquad \qquad \qquad + \sum_{l \in \mathcal{C}} P_{\mathrm{s}}^{(1)} (l | i) \Pr (C_{2N+1} = j | C_{N+1} = l) .
\end{align*}
Consequently, the conditional probability of being able to decode for the first time at attempt two with final state $j$ and under initial state $i$ is
\begin{equation*}
\begin{split}
P_{\mathrm{s}}^{(2)} (j | i)
&= \Pr (C_{2N+1} = j | C_1 = i) - P_{\mathrm{f}}^{(2)} (j | i) \\
&- \sum_{l \in \mathcal{C}} P_{\mathrm{s}}^{(1)} (l | i) \Pr (C_{2N+1} = j | C_{N+1} = l) .
\end{split}
\end{equation*}
Extending this procedure, we can compute the probability of a decoding success at attempt~$s$ with final state~$j$, given initial state~$i$,
\begin{equation*}
\begin{split}
P_{\mathrm{s}}^{(s)} (j | i)
&= \Pr \left( C_{sN+1} = j | C_1 = i \right) - P_{\mathrm{f}}^{(s)} (j | i) \\
&- \sum_{r=1}^{s-1} \sum_{l \in \mathcal{C}} P_{\mathrm{s}}^{(r)} (l | i ) \Pr \left(  C_{sN+1} = j  | C_{rN+1} = l \right) .
\end{split}
\end{equation*}
This methodology provides a recursive and efficient way to compute the probabilities that, under hybrid ARQ, a system takes exactly $s$ coded chunks to decode the original message.
As in Section \ref{section:AutomaticRepeatRequest}, we intend to compute the matrix generating function of $T$, the time spent in the first level of the reduced Markov chain.

Consider the aforementioned hybrid ARQ scheme with depth equal to $a$.
When there is a decoding failure at attempt $a$, the hybrid ARQ system has a few potential options.
The system can discard previously received symbols altogether and start the process anew.
Alternatively, the transmitter can re-encode the data segment and the information in previously received symbols can be used as side information during the decoding process.
No matter what the exact strategy is, the queue occupancy of a hybrid ARQ system can always be lower and upper bounded.
\begin{itemize}

\item \emph{Lower bound}: In this mode, the decoding of a message always succeeds by the $a$th attempt.
We call this the optimistic system.
Let $\check{T}$ denote the time spent in the first level of the reduced Markov chain associated with this system.

\item \emph{Upper bound}: In this mode, whenever decoding fails at the $a$th attempt, previously received symbols are discarded altogether and the process starts anew.
We call this the pessimistic view.
Let $\hat{T}$ denote the time spent in the first level of the reduced Markov chain of this system.

\end{itemize}
In essence, $\check{T}$ and $\hat{T}$ are Markov times that provide lower and upper bounds on $T$, the true stopping time of the hybrid ARQ decoding process.
These strategies jointly produce a near complete characterization of the behavior of hybrid ARQ systems.
We turn to the specifics of the proposed approaches below.

As mentioned above, an optimistic bound (lower bound) on $T$ can be derived using
\begin{equation*}
\begin{split}
\hat{P}_{\mathrm{s}}^{(a)} (j | i)
&= \Pr \left(  C_{aN+1} = j  | C_1 = i \right) \\
&- \sum_{r=1}^{a-1}\sum_{l \in \mathcal{C}} P_{\mathrm{s}}^{(r)} (l | i)
\Pr \left(  C_{aN+1} = j  | C_{rN+1} = l \right) ,
\end{split}
\end{equation*}
instead of $P_{\mathrm{s}}^{(a)} (j | i)$, by assuming that the decoding always succeeds by the $a$th attempt.
This bound holds irrespective of how the system handles failures at attempt~$a$.
We define the optimistic matrix generating function $\mathbf{G}_{\check{T}} (z) = \mathbf{G}_{\min{ \{T, a \}}} (z)$ entrywise by
\begin{equation*}
\begin{split}
\left[ \mathbf{G}_{\check{T}} (z) \right]_{ij}
= \sum_{r=1}^{a-1}  P_{\mathrm{s}}^{(r)} (j | i) z^r +  \hat{P}_{\mathrm{s}}^{(a)} (j | i) z^a.
\end{split}
\end{equation*}

The pessimistic matrix generating function $\mathbf{G}_{\hat{T}} (z)$ can be derived in two steps.
First, consider the matrix generating function
\begin{equation*}
\left[ \mathbf{G}_{\tilde{T}} (z) \right]_{ij}
= \sum_{r=1}^{a}  P_{\mathrm{s}}^{(r)} (j | i) z^r
\end{equation*}
Then, under the assumption that information is discarded when the $a$ decoding attempts have failed, we get
\begin{equation*}
\begin{split}
\mathbf{G}_{\hat{T}} (z)
&= \sum_{t = 0}^{\infty} z^{at}  \left( \mathbf{P}_{\mathrm{f}}^{(a)} \right)^t \mathbf{G}_{\tilde{T}} (z)
= \left( \mathbf{I} - z^a  \mathbf{P}_{\mathrm{f}}^{(a)} \right)^{-1} \mathbf{G}_{\tilde{T}} (z) .
\end{split}
\end{equation*}
Above, the matrix $\mathbf{P}_{\mathrm{f}}^{(a)}$ is defined entrywise as
\begin{equation*}
\left[ \mathbf{P}_{\mathrm{f}}^{(a)} \right]_{ij} = P_{\mathrm{f}}^{(a)}(j | i) .
\end{equation*}
We will return to these bounds and their application in Section~\ref{section:NumericalResults}.

\subsection{Hitting Time to an Empty Buffer}

We can build upon the matrix generating function of $T$ to obtain the distribution of $H_0$.
The basic insights behind this characterization are that the sojourn time at any level is finite almost surely and generating matrices can account for conditional independence.

\begin{theorem} \label{theorem:GeneratingFunctionH0}
The ordinary generating function of $H_0$, the first-passage time to an empty queue, is given by
\begin{equation} \label{equation:GeneratingFunctionEmptyQueue}
\begin{split}
G_{H_0} (z) = \mathrm{E} \left[ z^{H_q}  \right]
= \pi_0 \left( \mathbf{G}_{T}(z) \right)^m
\begin{bmatrix} 1 \\ \vdots \\ 1 \end{bmatrix}
\end{split}
\end{equation}
where $\pi_0$ is the channel state probability vector at time zero.
\end{theorem}
\begin{IEEEproof}
This expression for $G_{H_0}(z)$ can be obtained from an application of mathematical induction, which proceeds backward in time.
The first step consists in showing that the hypothesis holds for the base case, the sojourn time at level~$m$,
\begin{equation*}
\begin{split}
&\left[ \pi_0 \mathbf{G}_{T_m} (z) \right]_j
= \sum_{i = 1}^k \left[ \mathbf{G}_{T_m} (z) \right]_{ij} \Pr (C_1 = i) \\
&= \sum_{i = 1}^k
\mathrm{E} \left[ z^{T_m} \mathbf{1}_{ \left\{ C_{NT_m+1} = j \right\} } \middle| C_1 = i \right] 
\Pr (C_1 = i) \\
&= \mathrm{E} \left[ z^{T_m} \mathbf{1}_{ \left\{ C_{NT_m+1} = j \right\} } \right]
= \mathrm{E} \left[ z^{H_{m-1}} \mathbf{1}_{ \left\{ C_{NH_{m-1}+1} = j \right\} } \right]
\end{split}
\end{equation*}
where we have used the fact that $H_{m-1} = T_m$.
Thus, we gather that
\begin{equation*}
\llbracket z^t \rrbracket \left[ \pi_0 \mathbf{G}_{T_m} (z) \right]_j
= \Pr \left( H_{m-1} = t, C_{NH_{m-1}+1} = j \right) .
\end{equation*}
We continue with the inductive step in a similar manner.
Suppose that the hypothesis is true for a certain integer $q$ where $0 < q < m$; that is,
\begin{equation*}
\begin{split}
\left[ \pi_0 \mathbf{G}_{H_q} (z) \right]_j
&= \mathrm{E} \left[ z^{H_{q}} \mathbf{1}_{ \{ C_{NH_q+1} = j \} } \right] \\
&= \left[ \pi_0 \mathbf{G}_{T_m}(z) \cdots \mathbf{G}_{T_{q+1}}(z) \right]_j .
\end{split}
\end{equation*}
Then, we can write
\begin{equation*}
\begin{split}
&\mathrm{E} \left[ z^{H_{q-1}} \mathbf{1}_{ \{ C_{NH_{q-1}+1} = j \} } \right]
= \mathrm{E} \left[ z^{H_q + T_q} \mathbf{1}_{ \{ C_{NH_{q-1}+1} = j \} } \right] \\
&= \sum_{i=1}^k \mathrm{E} \left[ z^{H_q + T_q}
\mathbf{1}_{ \{ C_{NH_{q-1}+1} = j \} } \middle| C_{NH_q+1} = i \right] \times \\ 
&\qquad \Pr (C_{NH_q+1} = i) \\
&= \sum_{i=1}^k \mathrm{E} \left[ z^{H_{q}} \mathbf{1}_{ \{ C_{NH_q+1} = i \} } \right] \times  \\
&\qquad \mathrm{E} \left[ z^{T_q} \mathbf{1}_{ \{ C_{NH_{q-1}+1} = j \} } \middle| C_{NH_q+1} = i \right] \\
&= \sum_{i=1}^k
\left[ \pi_0 \mathbf{G}_{T_m}(z) \cdots \mathbf{G}_{T_{q+1}}(z) \right]_i
\left[ \mathbf{G}_{T_q} (z) \right]_{ij} \\
&= \left[ \pi_0 \mathbf{G}_{T_m} (z) \cdots \mathbf{G}_{T_q}(z) \right]_j
= \left[ \pi_0 \mathbf{G}_{H_{q-1}} (z) \right]_j .
\end{split}
\end{equation*}
That is, the hypothesis is also true for $q-1$.
We note that the third equality follows from the conditional independence of our quasi-birth-death Markov process.
In our problem, we have $\mathbf{G}_{T_q} (z) = \mathbf{G}_T (z)$ for all $q \in \{ 1, \dots, m \}$.
Since this expression holds for any $\pi_0$, we conclude that $\mathbf{G}_{H_0} (z) = \left( \mathbf{G}_T (z) \right)^m$ and, as a consequence,
\begin{equation*}
\llbracket z^t \rrbracket \left[ \pi_0 \left( \mathbf{G}_T (z) \right)^m \right]_j
= \Pr \left( H_{0} = t, C_{NH_0+1} = j \right) .
\end{equation*}
Summing over all the possible end states, we recover the expression for $G_{H_0} (z)$ given in \eqref{equation:GeneratingFunctionEmptyQueue}.
\end{IEEEproof}
To differentiate among possible initial conditions, it will become useful to write the first-passage time to an empty queue with an initial buffer size of $m$ segments as $H_0^{(m)}$.

\section{Large Deviation Analysis}
\label{section:LDP}

As seen in the previous section, it is possible to evaluate the exact distribution of $H_{0}^{(m)}$.
This facilitates the selection of parameters to optimize overall performance.
However, this process becomes cumbersome for large buffer sizes.
In such circumstances, analyzing the large deviations governing the system offers a new direction to derive meaningful guidelines for resource allocation and parameter tuning.
Below, we study two types of aberrations under the ARQ scheme: deviations in the average transmission time and the mean service rate.
We note that, although large deviations can be studied under hybrid ARQ, this latter scenario is somewhat tedious and it offers limited additional insights.
Hence we restrict our attention to the ARQ scheme. 
We begin with the average transmission time; that is, the normalized first-passage time to an empty queue.

\subsection{Normalized First-Passage Time}
\label{subsection:LDonHitting}

Again, suppose that the transmit buffer contains exactly $m$ segments at the onset of the communication process.
We are interested in the large deviations associated with the sequence of random variables specified by
\begin{equation*}
Y_m = \frac{1}{m} H_0^{(m)}
= \frac{1}{m} \sum_{q=1}^m T_q \quad m = 1, 2, \ldots
\end{equation*}
The logarithmic moment generating function for $Y_m$ is
\begin{equation*}
\begin{split}
\Lambda_m (\lambda) &= \log \mathrm{E} \left[ e^{\lambda Y_m} \right]
= \log \mathrm{E} \left[ e^{\lambda H_0^{(m)}/m} \right] \\
&= \log G_{H_0}^{(m)} \left( e^{{\lambda}/{m}} \right) .
\end{split}
\end{equation*}
The existence of limits of properly scaled logarithmic moment generating functions suggests that $\{ Y_m \}$ may satisfy a large deviation principle~\cite{Dembo0387984062}.
In particular, consider the following asymptotic regime
\begin{equation} \label{equation:MomentGeneratingFunction}
\begin{split}
\Lambda (\lambda) &= \lim_{m \rightarrow \infty} \frac{1}{m} \Lambda_m (m \lambda)
= \lim_{m \rightarrow \infty} \frac{1}{m} \log G_{H_0}^{(m)} \left( e^{\lambda} \right) \\
&= \lim_{m \rightarrow \infty} \frac{1}{m} \log \left( \pi_0 \left(
\mathbf{G}_{T} \left( e^{\lambda} \right) \right)^m \mathbf{1} \right) .
\end{split}
\end{equation}
A few observations concerning $\Lambda (\lambda)$ are in order.
In view of Lemma~\ref{lemma:BitErasureGeneratingMatrix} and for $z = e^{\lambda}$,
\begin{equation*}
\mathbf{G}_T \left( e^{\lambda} \right)
= \left( \sum_{t=0}^{\infty} \mathbf{K}^t e^{t \lambda} \right) \mathbf{M} e^{\lambda}
\end{equation*}
is a non-negative matrix over the extended real numbers.
In fact, this matrix possesses additional properties which are summarized below.
Again, let $\varrho(\cdot)$ denote the spectral radius of its matrix argument.
\begin{lemma} \label{lemma:RadiusOfConvergenceOfGT}
  If $T$ is finite almost surely, the matrix generator $\mathbf{G}_T\left(e^{\lambda}\right)$ exists as a non-negative real matrix  if and only if $\lambda < -\log \varrho(\mathbf{K})$.
In particular, when $\lambda \geq - \log \varrho(\mathbf{K})$, one or more entries of $\mathbf{G}_T(e^{\lambda})$ will be infinite.
\end{lemma}
\begin{IEEEproof}
See Appendix \ref{appendix:RadiusOfConvergenceOfGT}.
\end{IEEEproof}

Another important quantity is the spectral radius of $\mathbf{K}$, which is related to the support of $\mathbf{G}_T\left(e^{\lambda}\right)$ as seen in Lemma~\ref{lemma:RadiusOfConvergenceOfGT}.

\begin{corollary} \label{corollary:SpectralRadiusOfK}
  If $T$ is finite almost surely, then $\varrho(\mathbf{K})<1$.
\end{corollary}
\begin{IEEEproof}
See Appendix \ref{appendix:SpectralRadiusOfK}.
\end{IEEEproof}

Under Assumption \ref{assumption:PrimitiveMarkovChannel} and for any non-trivial coding scheme, $T$ is finite almost surely, thus the hypotheses of Lemma~\ref{lemma:RadiusOfConvergenceOfGT} and Corollary~\ref{corollary:SpectralRadiusOfK} are satisfied.
A sufficient condition to ensure the existence of a large deviation principle for the average transmission time is that the Markov process $\{ U_t \}$ sampled at departure events $\{ H_q \}$ is irreducible.
This guarantees that the states of the corresponding jump chain form a unique recurrent class.
Formally, we postulate the following condition.

\begin{assumption} \label{assumption:IrreducibilityOfGT}
The matrix $(\mathbf{I}-\mathbf{K})^{-1}\mathbf{M}$ is irreducible.
\end{assumption}

We note that, strictly speaking, this is not a necessary condition.
Having a unique communicating class and, possibly, transient states in the jump chain will also work.
However, this more encompassing setting leads to extra bookkeeping, which unnecessarily clouds some of the underlying concepts. 
Furthermore, all the practical systems we wish to study fulfill the requirements of Assumption~\ref{assumption:IrreducibilityOfGT}.
As such, we take it for granted from this point forward.
Under this assumption, the matrix $\mathbf{G}_T\left(e^{\lambda}\right)$ is irreducible for any $\lambda < - \log \varrho(\mathbf{K})$ and, hence, the Perron-Frobenius theorem applies \cite{Horn0521386322,Dembo0387984062}.
This leads to the following result.

\begin{proposition} \label{proposition:AsymptoticMGF}
Under Assumption~\ref{assumption:IrreducibilityOfGT}, the limiting moment generating function defined in \eqref{equation:MomentGeneratingFunction} exists as an extended real number for every $\lambda \in \mathbb{R}$, with
\begin{equation*}
\Lambda (\lambda)
= \begin{cases}
\varrho \left( \left( \mathbf{I} - \mathbf{K} e^{\lambda} \right)^{-1}
\mathbf{M} e^{\lambda} \right) & \lambda < - \log \varrho (\mathbf{K}) \\
\infty & \text{otherwise} . \end{cases}
\end{equation*}
\end{proposition}
\begin{IEEEproof}
  See Appendix \ref{appendix:AsymptoticMGF}.
\end{IEEEproof}

Using matrix norms, it can be shown that $\mathbf{G}_T \left( e^{\lambda} \right)$ is differentiable entrywise over the interval $\lambda < - \log \varrho (\mathbf{K})$.
Since $\Lambda (\lambda)$ is an isolated root of the characteristic function of matrix $\mathbf{G}_T \left( e^{\lambda} \right)$, we deduce that it is positive, finite and differentiable with respect to $\lambda$ (see, e.g., \cite[Th.~11.5.1]{Lancaster0124355609}, \cite[p.~75]{Dembo0387984062}).
Corollary~\ref{corollary:SpectralRadiusOfK} asserts that $\varrho(\mathbf{K})<1$, which implies that $\Lambda (0)$ is finite.
In view of the discussion above, we conclude that the origin is in the interior of $\{ \lambda \in \mathbb{R} : \Lambda(\lambda) < \infty \}$.
Consequently, $\Lambda (\lambda)$ is essentially smooth and the G\"artner-Ellis theorem applies~\cite{Dembo0387984062}, thereby establishing the desired result.

\begin{theorem} \label{theorem:HittingLDP}
Suppose $\left\{ Y_m = \frac{1}{m} \sum_{q=1}^m T_q \right\}$ is the empirical mean sojourn time per level.
For every $x \in \mathbb{R}$, consider the Fenchel-Legendre transform
\begin{equation}
\Lambda^* (x) = \sup_{\lambda \in \mathbb{R}} \left\{ \lambda x
- \log \varrho \left( \mathbf{G}_T \left( e^{\lambda} \right) \right) \right\} .
\end{equation}
The empirical mean $Y_m$ satisfies the large deviation principle with the convex, good rate function $\Lambda^* (\cdot)$.
That is, for any set $\Gamma \subseteq \mathbb{R}$ and any initial state $c \in \mathcal{C}$,
\begin{equation*}
\begin{split}
- \inf_{x \in \Gamma^{o}} \Lambda^* (x)
&\leq \liminf_{m \rightarrow \infty} \frac{1}{m} \log \Pr (Y_m \in \Gamma) \\
&\leq \limsup_{m \rightarrow \infty} \frac{1}{m} \log \Pr (Y_m \in \Gamma)
\leq - \inf_{x \in \bar{\Gamma}} \Lambda^* (x) ,
\end{split}
\end{equation*}
where $\Gamma^{o}$ and $\bar{\Gamma}$ denote the interior and closure of the set $\Gamma$, respectively.
\end{theorem}

\begin{example}
For the Gilbert-Elliott channel shown in Fig.~\ref{figure:FiniteStateChannel}, it is possible to obtain a closed-form expression for the spectral radius of $\mathbf{G}_T \left( e^{\lambda} \right)$.
Specifically, we can write the characteristic polynomial of $\mathbf{G}_T \left( e^{\lambda} \right)$ as
\begin{equation*}
\begin{split}
\operatorname{det} \left( \gamma \mathbf{I}
- \mathbf{G}_T \left( e^{\lambda} \right) \right)
&= \operatorname{det} \left( \gamma \mathbf{I}
- \left( \mathbf{I} - \mathbf{K} e^{\lambda} \right)^{-1}
\mathbf{M} e^{\lambda} \right) \\
&= \frac{ \operatorname{det} \left( \gamma \mathbf{I}
- \gamma \mathbf{K} e^{\lambda} - \mathbf{M} e^{\lambda} \right) }
{ \operatorname{det} \left( \mathbf{I} - \mathbf{K} e^{\lambda} \right) } .
\end{split}
\end{equation*}
We note that the numerator is a quadratic equation in $\gamma$ and the denominator is a constant.
It is therefore possible to find parametric expressions for the two roots of $\operatorname{det} \left( \gamma \mathbf{I} - \mathbf{G}_T \left( e^{\lambda} \right) \right)$.
Taking the maximum of the absolute values of these two roots yields an explicit, albeit convoluted, expression for the spectral radius of $\mathbf{G}_T \left( e^{\lambda} \right)$.
As such, $\Lambda^* (\cdot)$ can be obtained efficiently.
\end{example}

\subsection{Empirical Mean Service}
\label{subsection:LDonService}

We turn to the second type of aberrations we wish to study: deviations in the empirical mean service rate,
\begin{equation*}
Z_s = \frac{1}{s} \sum_{t=1}^{s} D_t .
\end{equation*}
We note that $\{ D_s \}$ is not a Markov process.
However, $D_s$ is a (trivial) deterministic function of $V_s = (C_{(s+1)N+1}, D_s)$.
Since $\{ V_s \}$ is a Markov process, we can apply general results on the large deviation principle of additive functionals of Markov chains.
To leverage these results, we first impose an ordering on the state space $\mathcal{V} = \mathcal{C} \times \{ 0, 1 \}$.
Recall that $| \mathcal{C} | = k$; a natural ordering for this state space is to associate integer $v = (dk + i)$ with state $(i, d)$.
Using this ordering, the transition probability matrix $\mathbf{\Pi}$ for the augmented process $\{V_s \}$ is given by
\begin{equation*}
\left[ \mathbf{\Pi} \right]_{v_1,v_2} = \pi (v_1, v_2), \;\;  v_1, v_2 \in  \{1, \ldots, 2 k\}
\end{equation*}
where $\pi (v_1, v_2)$ is the probability of jumping to state $v_2$, conditioned on starting from $v_1$.

\begin{assumption} \label{assumption:IrreducibilityOfPi}
The matrix $\mathbf{\Pi}$ is irreducible.
\end{assumption}

This assumption is similar in spirit to Assumption~\ref{assumption:IrreducibilityOfGT}.
Yet the large deviation principle on the empirical service can be derived under weaker conditions.
In particular, it suffices to show that $\mathbf{K} + \mathbf{M}$ is irreducible, a requirement that is easily met.
We stress that $\mathbf{K} + \mathbf{M}$ is equal to $\mathbf{B}^N$, and the latter matrix is itself irreducible by Assumption~\ref{assumption:PrimitiveMarkovChannel}.

\begin{theorem}[\cite{Dembo0387984062}] \label{theorem:MarkovLDP}
Let $\{ V_s \}$ be a finite-state Markov chain possessing an irreducible transition matrix $\mathbf{\Pi}$.
For every $x \in \mathbb{R}$, define
\begin{equation}
I(x) = \sup_{\lambda \in \mathbb{R}}
\left\{ \lambda x - \log \varrho \left( \mathbf{\Pi}_{\lambda} \right) \right\}
\end{equation}
where $\mathbf{\Pi}_{\lambda}$ is a nonnegative matrix whose elements are
\begin{equation*}
\pi_{\lambda} \left( v_1, v_2 \right) = \pi(v_1, v_2) e^{\lambda d_2} \quad
v_1, v_2 \in \{1, \ldots, 2k \} .
\end{equation*}
Then, the empirical mean $Z_s$ satisfies the large deviation principle with the convex good rate function $I(\cdot)$.
Explicitly, for any set $\Gamma \subseteq \mathbb{R}$, and any initial state $v \in \mathcal{V}$,
\begin{equation*}
\begin{split}
- \inf_{x \in \Gamma^{o}} I(x)
&\leq \liminf_{s \rightarrow \infty} \frac{1}{s} \log P_v^{\pi} (Z_s \in \Gamma) \\
&\leq \limsup_{s \rightarrow \infty} \frac{1}{s} \log P_v^{\pi} (Z_s \in \Gamma)
\leq - \inf_{x \in \bar{\Gamma}} I(x)
\end{split}
\end{equation*}
where $P_v^{\pi}$ denotes the Markov probability measure induced by transition probability $\mathbf{\Pi}$ and initial state $v \in \mathcal{V}$, i.e.,
\begin{equation*}
P_v^{\pi} (V_1 = v_1, \ldots, V_s = v_s)
= \pi (v, v_1) \prod_{t=1}^{s-1} \pi (v_t, v_{t+1}) .
\end{equation*}
\end{theorem}

Expressions for the transition probabilities used in this theorem appear in \eqref{equation:AugmentedTransitionProbabilities}.
We note that
\begin{equation*}
\begin{split}
\Pr & \left( V_{s+1} = (j, d_2) | V_s = (i, d_1) \right) \\
&= \Pr \left( V_{s+1} = (j, d_2) | C_{(s+1)N+1} = i \right) ;
\end{split}
\end{equation*}
this induces a repetitive structure in matrix $\mathbf{\Pi}$.
The nonnegative matrix $\mathbf{\Pi}_{\lambda}$ associated with every $\lambda \in \mathbb{R}$ can then be written explicitly as
\begin{equation} \label{equation:MatrixPiLambda}
\mathbf{\Pi}_{\lambda} = \begin{bmatrix}
\kappa_{11}	& \cdots & \kappa_{1k} & 
\mu_{11} e^{\lambda} & \cdots & \mu_{1k} e^{\lambda} \\
\vdots & \ddots & \vdots & \vdots & \ddots & \vdots \\
\kappa_{k1} & \cdots & \kappa_{kk} & 
\mu_{k1} e^{\lambda} & \cdots & \mu_{kk} e^{\lambda} \\
\kappa_{11} & \cdots & \kappa_{1k} & 
\mu_{11} e^{\lambda} & \cdots & \mu_{1k} e^{\lambda}\\
\vdots & \ddots & \vdots & \vdots & \ddots & \vdots \\
\kappa_{k1} & \cdots & \kappa_{kk} & 
\mu_{k1} e^{\lambda} & \cdots & \mu_{kk} e^{\lambda} 
\end{bmatrix} .
\end{equation}
We can rewrite $\mathbf{\Pi}_{\lambda}$ by taking advantage of its block structure,
\begin{equation*}
\mathbf{\Pi}_{\lambda} = \begin{bmatrix}
\mathbf{K} & \mathbf{M} e^{\lambda}  \\
\mathbf{K} & \mathbf{M} e^{\lambda} 
\end{bmatrix} .
\end{equation*}
The pertinent eigenvalues are the roots of the characteristic polynomial of $\mathbf{\Pi}_{\lambda}$.
Using properties of matrix determinant and the commutative properties of some of the blocks, we can express this polynomial as
\begin{equation*}
\begin{split}
\operatorname{det} \left( \gamma \mathbf{I} - \mathbf{\Pi}_{\lambda} \right)
&= \operatorname{det} \left( (\gamma \mathbf{I} - \mathbf{M} e^{\lambda})
(\gamma \mathbf{I} - \mathbf{K}) - \mathbf{M} \mathbf{K} e^{\lambda} \right) \\
&= \operatorname{det} \left( (\gamma \mathbf{I} - \mathbf{K})
(\gamma \mathbf{I} - \mathbf{M} e^{\lambda})
- \mathbf{K} \mathbf{M} e^{\lambda} \right) \\
&= \operatorname{det} \left( \gamma^2 \mathbf{I} - \gamma \mathbf{K} - \gamma \mathbf{M} e^{\lambda} \right) .
\end{split}
\end{equation*}

Collectively, Theorem~\ref{theorem:MarkovLDP} and the matrix defined in \eqref{equation:MatrixPiLambda} provide an algorithmic workflow for the computation of the good rate function associated with the empirical means $\{ Z_s \}$.
We follow this discussion with an example based on a two-state channel with memory.

\begin{example} \label{example:GilbertElliottEMS}
Once again, consider a Gilbert-Elliott erasure channel with $\mathcal{C} = \{ 1, 2 \}$.
An advantage in studying this rudimentary model is that it admits a simple, closed-form characterization.
The dimension of the state space in this case is $| \mathcal{V} | = 4$.
Using the commutative block structure discussed above, the determinant of $(\gamma \mathbf{I} - \mathbf{\Pi}_{\lambda})$ reduces to
\begin{equation*}
\begin{split}
\operatorname{det} & \left( \gamma \mathbf{I} - \mathbf{\Pi}_{\lambda} \right)
= \operatorname{det} \left( \gamma^2 \mathbf{I} - \gamma \mathbf{K}
- \gamma \mathbf{M} e^{\lambda} \right) \\
&= \gamma^2 \operatorname{det} \left( \begin{bmatrix}
\gamma - \kappa_{11} - \mu_{11} e^{\lambda} & - \kappa_{12} - \mu_{12} e^{\lambda} \\
- \kappa_{21} - \mu_{21} e^{\lambda} & \gamma - \kappa_{22} - \mu_{22} e^{\lambda}
\end{bmatrix} \right) .
\end{split}
\end{equation*}
By inspection, we see that the spectral radius of $\mathbf{\Pi}_{\lambda}$ is the largest root of the quadratic equation
\begin{equation*}
\begin{split}
\gamma^2 &- \gamma (\kappa_{11} + \kappa_{22} + (\mu_{11} + \mu_{22}) e^{\lambda}) \\
&+ (\kappa_{11} + \mu_{11} e^{\lambda})(\kappa_{22} + \mu_{22} e^{\lambda}) \\
&- (\kappa_{12} + \mu_{12} e^{\lambda})(\kappa_{21} + \mu_{21} e^{\lambda})  = 0 .
\end{split}
\end{equation*}
For fixed parameters, this dominating root can be computed using the celebrated quadratic formula.
We will revisit this example in Section~\ref{section:NumericalResults}.
\end{example}

\subsection{Relation between $\Lambda^{*}(\cdot)$ and $I(\cdot)$}
\label{subsection:relatingLDPs}

The two rate functions introduced above, $\Lambda^* (\cdot)$ and $I (\cdot)$, characterize the large deviation principles for the mean transmission time and average service rate, respectively.
Since the processes $\{ T_q \}$ and $\{ D_s \}$ are closely related, one can presume that their governing rate functions are somehow linked.
A key insight in understanding this relation is to realize that the following events are equivalent:
for any positive integers $m$ and $n$,
\begin{equation} \label{equation:EquivalentEvents}
\{ T_1 + \cdots + T_m >n \} = \{ D_1 + \cdots + D_n < m \} .
\end{equation}
In words, the first event occurs whenever more than $n$ attempts are required to successfully deliver $m$ packets, while the second event states that fewer than $m$ packet transmissions have been successful within the first $n$ attempts. 
Using this relationship and scaling arguments, one can establish our next proposition which substantiates the existence of a strong connection between the two rate functions.

\begin{proposition}
\label{proposition:RelationILambda}
If the rate functions $\Lambda^*(\cdot)$ and $I(\cdot)$ are finite in the open intervals $(1,\infty)$ and $(0,1)$, respectively, then they satisfy
\begin{equation*}
I(x) = x \Lambda^* \left( \frac{1}{x} \right)
\end{equation*}
for $x \in (0, 1)$.
\end{proposition}
\begin{IEEEproof}
See Appendix \ref{appendix:RelationILambda}.
\end{IEEEproof}

\section{Performance Evaluation}
\label{section:performance}

Thus far, we have devoted much attention to developing a thorough understanding of $H_0$ and, in particular, its generating function.
In this section, we apply the results of Theorem~\ref{theorem:GeneratingFunctionH0} and we derive a number of pertinent performance criteria with practical significance.

First, recall that $\llbracket z^t \rrbracket G_{H_0} (z) = \Pr (H_0 = t)$.
Accordingly, the probability that the queue fails to drain within $\tau$ time units is equal to
\begin{equation*}
\Pr (H_0 > \tau) = 1 - \sum_{t=0}^{\lfloor \tau \rfloor} 
\llbracket z^t \rrbracket G_{H_0} (z) .
\end{equation*}
Moreover, the average time required to empty the queue is obtained by differentiating the moment generating function of $H_0$ and then taking the limit as $z$ approaches one,
\begin{equation*}
\mathrm{E} \left[ H_0 \right] = \lim_{z \uparrow 1} \frac{d}{dz} G_{H_0} (z) .
\end{equation*}

Alternatively, using Chernoff inequalities, it is possible to upper bound the probability of a deviation event in a computationally efficient manner.
The equation
\begin{equation*}
\Pr (H_0 > \tau)
\leq e^{-\lambda \tau} \mathrm{E} \left[ e^{\lambda H_0} \right]
= e^{-\lambda \tau} G_{H_0} \left( e^{\lambda} \right)
\end{equation*}
holds for any $\lambda > 0$.
The optimal bound derived from this collection of inequalities is sometimes expressed in logarithmic form,
\begin{equation*}
\log \Pr (H_0 > \tau)
\leq - \sup_{\lambda > 0}
\left\{ \lambda \tau - \log \left( G_{H_0} \left( e^{\lambda} \right) \right) \right\} .
\end{equation*}
The large deviation principle on $H_0$ derived in Section~\ref{section:LDP} confirms that, under mild conditions, this latter bound is asymptotically tight.

It may be instructive to stress that $H_0$, the first-passage time introduced in \eqref{equation:HittingTimes}, is defined in terms of codeword transmission attempts.
That is, $H_0$ represents the cumulative number of codewords sent by the source until the queue empties out completely.
Such a metric poses no issue when comparing systems of identical block lengths.
However, when assessing the performance of candidate implementations with different block lengths, a more careful interpretation of the results becomes necessary.
This subtlety arises because of the mismatch in indexing between the evolution of the queue and the number of channel uses.
For a fair evaluation of potential candidates, hitting times should be scaled to portray their evolution according to a common clock, that of the channel process.

Define random variable $\tilde{H}_0$ by
\begin{equation*}
\tilde{H}_0 = N H_0 ,
\end{equation*}
where $N$ designates the block length associated with the underlying implementation.
Then, $\tilde{H}_0$ denotes the number of channel uses necessary to empty out the queue, and it can therefore be employed to provide a uniform measure of performance.
While it is straightforward to extend our performance criteria to $\tilde{H}_0$ through the relation
\begin{equation*}
\Pr (H_0 > \tau) = \Pr \left( \tilde{H}_0 > \frac{\tau}{N} \right) ,
\end{equation*}
it is essential to apply this transformation when comparing systems with different block lengths.

A similar scaling is needed when comparing the large deviations of systems with different parameters.
A proper scaling for the fair comparison of mean sojourn times can be expressed in terms of channel uses per information bit, 
\begin{equation*}
\tilde{Y}_{\ell} = \frac{1}{\ell}
N H_0^{\left( \lceil \ell/K \rceil \right)} .
\end{equation*}
This leads to the following asymptotic regime
\begin{equation*}
\begin{split}
\lim_{\ell \rightarrow \infty} & \frac{1}{\ell} \log
\Pr \left( \tilde{Y}_{\ell} > \tau \right) \\
&= \frac{1}{K} \lim_{\ell \rightarrow \infty} \frac{1}{\lceil \ell / K \rceil}
\log \Pr \left( \frac{1}{\lceil \ell / K \rceil}
H_0^{\left( \lceil \ell / K \rceil \right)} > \frac{K}{N} \tau \right) \\
&= \frac{1}{K} \lim_{m \rightarrow \infty} \frac{1}{m}
\log \Pr \left( \frac{1}{m} H_0^{(m)} > \frac{K}{N} \tau \right) \\
&= - \frac{1}{K} \Lambda^* \left( \frac{K}{N} \tau \right)
\end{split}
\end{equation*}
where $\tau > \mathrm{E} \left[ \tilde{Y}_{\infty} \right]$.
Likewise, to account for discrepancies in design parameters, the empirical mean service can be expressed in terms of decoded bits per channel use,
\begin{equation*}
\tilde{Z}_n = \frac{1}{n} \sum_{t=1}^{\lfloor n/N \rfloor} K D_t .
\end{equation*}
The ensuing asymptotic regime becomes
\begin{equation*}
\begin{split}
\lim_{n \rightarrow \infty} & \frac{1}{n} \log \Pr \left( \tilde{Z}_n < \eta \right) \\
&= \frac{1}{N} \lim_{n \rightarrow \infty} \frac{1}{\lfloor n/N \rfloor}
\log \Pr \left( \frac{1}{\lfloor n/N \rfloor}
\sum_{t=1}^{\lfloor n/N \rfloor} D_t < \frac{N}{K} \eta \right) \\
&= \frac{1}{N} \lim_{s \rightarrow \infty} \frac{1}{s}
\log \Pr \left( \frac{1}{s}
\sum_{t=1}^{s} D_t < \frac{N}{K} \eta \right) \\
&= - \frac{1}{N} I \left( \frac{N}{K} \eta \right)
\end{split}
\end{equation*}
where $\eta < \mathrm{E} \left[ \tilde{Z}_{\infty} \right]$.
Collectively, these various modifications enables the comparison of competing implementations with different values for $K$ and $N$.

Another concern that comes into play when optimizing over  block length is the impact of the initial state of the system.
If the number of bits at the source is fixed at time zero, the scope of the optimal solution may be very narrow.
This is a situation akin to over-fitting in statistical modeling.
To provide a more robust characterization with widely applicable results and guidelines, it may be beneficial to assume that the number of bits in the queue at the onset of the transmission process is random, with a prescribed representative distribution.
In our numerical study, we circumvent some of these difficulties by assuming that the block length is fixed and the initial queue length is random.
The specifics of our investigation are detailed below.

\section{Numerical Analysis}
\label{section:NumericalResults}

In this section, we apply the methodology developed above to an illustrative example.
Physical parameters are selected to resemble an implementation of the global system for mobile communications (GSM).
Specifically, the block length is fixed at $N = 114$.
The information content per codeword, $K$, is a parameter to be optimized.
We model the wireless connection as a Gilbert-Elliott erasure channel, and we denote its transition probability matrix as
\begin{equation*}
\mathbf{B} = \begin{bmatrix} b_{11} & b_{12} \\
b_{21} & b_{22} \end{bmatrix} .
\end{equation*}
For simplicity, we assume that $\varepsilon_1 = 1$ and $\varepsilon_2 = 0$.
The probability of a bit erasure is set at twenty percent, which entails
\begin{equation*}
\frac{b_{21}}{b_{12} + b_{21}} = 0.2 .
\end{equation*}
For this elementary model, channel memory can be expressed unambiguously through the decay factor $(1 - b_{12} - b_{21})$, which is determined by the spectrum of the matrix.
A decay factor equal to zero is equivalent to a memoryless channel, while correlation increases as $(1 - b_{12} - b_{21})$ approaches one.
Except where specified otherwise, we employ a decay factor equal to $0.9$ in our numerical results.

We assume that $L$, the number of information bits contained at the source at time zero, is a random variable possessing a Gamma distribution with mean $2000$ and standard deviation $100$.
Randomizing the number of bits at the source partly alleviates the idiosyncratic effects associated with partitioning the queue content into segments of $K$ bits.
For a source buffer with $\ell$ information bits, the number of segments to be delivered is $\lceil \ell / K \rceil$ and, as such, a one-bit variation in $\ell$ can result in having an additional message to send.
Imposing a random distribution on the number of information bits at the source leads to a probability distribution on $M = \lceil L / K \rceil$.
This, in turn, yields smoother results.

\begin{figure}[bht]
\begin{center}
\setlength\tikzheight{5cm}
\setlength\tikzwidth{6.0cm}
\input{Figures/MeanHittingTime.tex}
\caption{This figure shows mean first-passage times as functions of $K$.
The block length employed in all cases is $N = 114$.
The underlying Gilbert-Elliott channel produces erasures with probability $0.20$, and it possesses a dominant decay factor of $(1 - b_{12} - b_{21}) = 0.9$.
The expected number of bits at the source at time zero is $2000$.
The upper and lower bounds for the hybrid ARQ scheme with a depth of $a=3$ are indistinguishable.}
\label{figure:MeanHittingTime}
\end{center}
\end{figure}
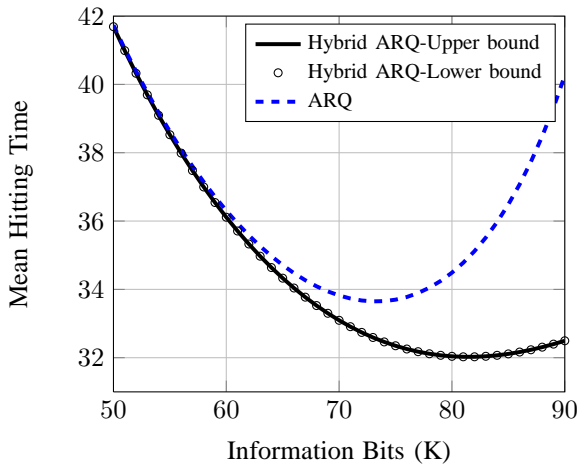

\begin{figure}[bth]
\begin{center}
\setlength\tikzheight{5cm} 
\setlength\tikzwidth{6.0cm} 
\input{Figures/VarHittingTime.tex}
\caption{This figure displays variances of the first-passage times to an empty queue as functions of $K$.
The parameters used in this numerical study are the same as those featured in Fig.~\ref{figure:MeanHittingTime}.
The variance for the hybrid ARQ scheme is calculated with the upper bound $\hat{T}$.}
\label{figure:VarHittingTime}
\end{center}
\end{figure}
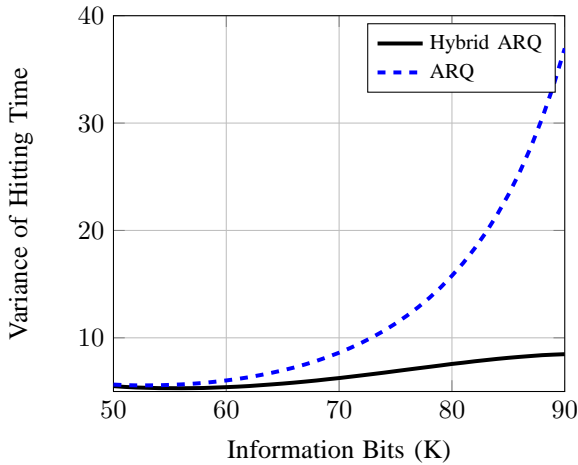

Figures~\ref{figure:MeanHittingTime} and \ref{figure:VarHittingTime} present the mean and variance of the first-passage times for the ARQ and hybrid ARQ schemes as functions of the number of information bits per codeword.
Varying the code rate affects both the expected value of the first-passage time and its variance.
A low code rate offers more protection against erasures and, accordingly, the resulting distribution of the hitting time to an empty queue is very narrow.
Increasing the code rate initially reduces the mean first-passage time, as every successful decoding attempt reveals more information bits.
However, a higher code rate also raises the probability of decoding failure.
Eventually, as the code rate is pushed further, decoding failures start to hamper the draining process and the mean first-passage time grows due to excessive repetition requests.
This effect is much more pronounced for standard ARQ.

The penalty in using a high code rate is less severe for the hybrid ARQ scheme because the failure recovery mechanism, which is based on incremental redundancy, adapts gracefully to channel conditions in this latter case.
For instance, when $K$ is very close to $N$, decoding under standard ARQ will fail nearly every time.
Contrastingly, the effective code rate drops rapidly with decoding failures under hybrid ARQ.
The robust profile of hybrid ARQ is a key property that underlies the popularity of this paradigm in practical systems.
In the current example, the upper and lower bounds derived for $\mathrm{E}[H_0]$ under the hybrid ARQ scheme are essentially indistinguishable, hinting at the fact that decoding failures are nearly nonexistent once three blocks are received.

Perhaps not too surprisingly, our numerical investigation suggests that the optimal code rate is somewhat impervious to initial queue conditions.
To examine the effects of the initial queue length, we employ the channel parameters described above and we modify the distribution on $L$.
For Gamma distributions with means $\mathrm{E}[L] \in \{ 500, 1000, 2000, 3000\}$ and standard deviation $100$, the optimal value of $K$ in terms of mean first-passage time is consistently equal to $73$ for standard ARQ and it remains fixed at $81$ for the hybrid variant.

\begin{figure}[bht]
\begin{center}
\setlength\tikzheight{5cm} 
\setlength\tikzwidth{6.0cm} 
\input{Figures/CrossingsCDF.tex}
\caption{The crossings of the cumulative distribution function $F_{H_0} (\cdot)$ offer conservative figures of merit for the operation of the queueing system.
In this example, the lines correspond to thresholds $p \in \{ 0.45, 0.95 \}$.}
\label{figure:CrossingsCDF}
\end{center}
\end{figure}
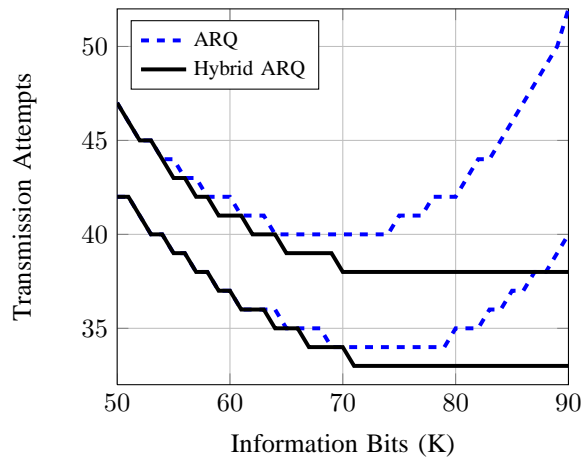

Using the methodology established thus far, it is possible to consider additional performance criteria.
For instance, we can analyze the crossings of the cumulative distribution function,
\begin{equation*}
h_p = \min_t \{  t | \Pr ( H_0 \leq t ) \geq p \} .
\end{equation*}
Fig.~\ref{figure:CrossingsCDF} plots the number of transmission attempts associated with threshold values $p \in \{ 0.45, 0.95 \}$.
We observe that the optimal value of $K$ decreases slightly when the crossing threshold $p$ approaches one.
In other words, when focusing on worst-case behavior, the system tends to favor a more conservative setting with extra protection against erasures.
This phenomenon offers another perspective on the tradeoff between expected behavior and its variations.

Next, we turn to the large deviations techniques developed in Section~\ref{section:LDP}.
As a reference, we consider a voice stream application.
In GSM, each speech frame of length 20~ms is encoded into a data segment of length 228.
The underlying physical layer has the ability to transmit one symbol every 40~$\mu$s.
If we approximate the maximum delay tolerance for one-way voice traffic to be 40~ms~\cite[p.~70]{TseFWC1111206}, then this requires $228$ bits to be transmitted within roughly $1000$ channel uses.
This constraint, in turn, necessitates a nominal rate on the order of 0.23~bits per channel use for link reliability.
We adopt this figure as a rough estimate for the needs of a voice stream in our numerical study.

\begin{figure}[t]
\begin{center}
\setlength\tikzheight{5cm} 
\setlength\tikzwidth{6.5cm} 
\input{Figures/RateFunctionService}
\caption{This figure plots good rate functions governing large deviations in the empirical mean service as functions of $K$, the number of information bits per codeword.
Given throughput threshold $\eta$, the optimal value of $K$ is the argument corresponding to the apex of the function.}
\label{figure:RateFunctionService}
\end{center}
\end{figure}
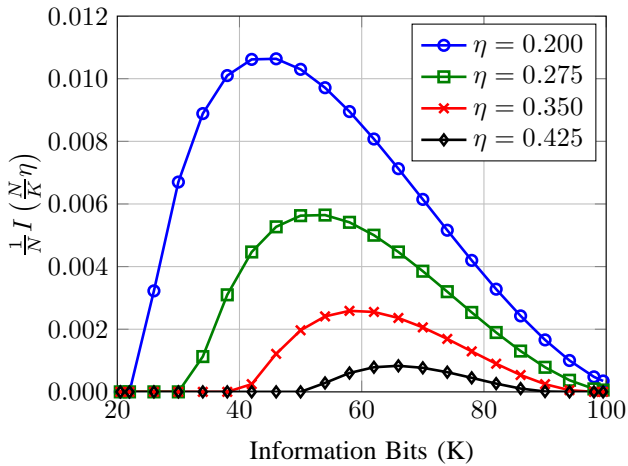

The maximum throughput that can be supported over the Gilbert-Elliott channel in our example is slightly above 0.5~bits per channel use.
Recall that threshold $\eta$ represents a minimum target requirement on the number of information bits per channel use that can be successfully decoded at the destination, in an asymptotic regime.
When $\eta < 0.5$, there exist values of $K$ for which the rate function
$\frac{1}{N} I \left( \frac{N}{K} \eta \right)$ is strictly positive;
this can be seen in Fig.~\ref{figure:RateFunctionService}.
These curves can be used to characterize the tension between quantization and failures to deliver media properly.
A high-quality stream, with a large $\eta$, will offer an enhanced viewer experience when transmitted adequately, but will necessarily be more prone to interruptions and failures, as exposed through the rate functions.
A low-bandwidth, low-quality stream on the other hand offers a better delivery profile with a smaller probability of failure.
However, the quality of the playback may not be satisfactory to the end user.
A proper selection of parameters for an adequate overall user experience can be made through the rate functions of Fig.~\ref{figure:RateFunctionService}.

Once $\eta$ is picked, the corresponding curve displays performance as a function of $K$.
For low code rates, the maximum achievable throughput is less than the service requirement and hence the rate function governing large deviations is zero.
At high code rates, performance is limited by the rise in the probability of decoding failure.
The system must then find the right balance between the frequency of failures and the payoff of a decoding success in terms of information bits.
The optimal value of $K$ for a specific threshold $\eta$ is given by the apex of its curve,
\begin{equation*}
K_Z^*(\eta) = \argmax_K \frac{1}{N} I \left( \frac{N}{K} \eta \right) .
\end{equation*}
It is interesting to note how conservative the optimal code rate becomes when the target service requirement is reduced.

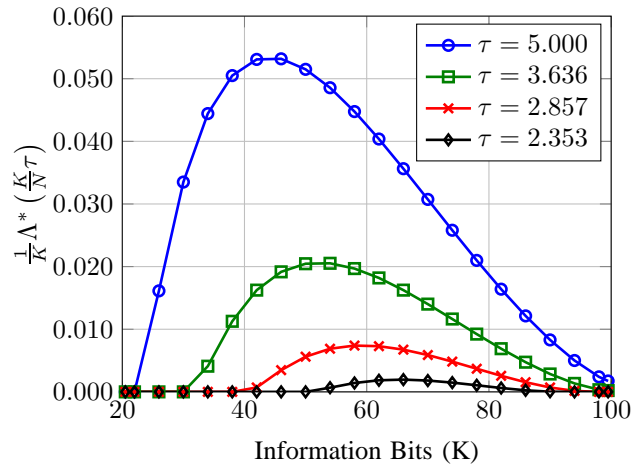
\begin{figure}[t]
\begin{center}
\setlength\tikzheight{5cm} 
\setlength\tikzwidth{6.5cm} 
\input{Figures/RateFunctionHitting}
\caption{This figure shows good rate functions governing large deviations in the mean sojourn time as functions of $K$.
The optimum code rate depends heavily on the deviation threshold of the mean sojourn time.}
\label{figure:RateFunctionHitting}
\end{center}
\end{figure}

The second type of rate functions introduced in Section~\ref{section:LDP} characterizes large deviations in the mean sojourn times, as shown in Fig.~\ref{figure:RateFunctionHitting}.
These curves can be employed to tradeoff playback quality and buffering times for streaming media.
More specifically, $\tau$ represents a limitation on the average number of channel uses employed to transmit one bit of information.
Of course, when a high-quality rendering is selected, the system must deliver a larger amount of data within the buffering window and, hence, the probability of delay violation becomes greater.
In this case, the optimal value of $K$ becomes
\begin{equation*}
K_Y^* (\tau) = \argmax_K \frac{1}{K} \Lambda^* \left( \frac{K}{N} \tau \right) .
\end{equation*}
The behavior of the system in terms of average sojourn time is closely related to the empirical mean service, holding a reciprocal relation.
We emphasize that the optimal code rates are equal, namely $K_Z^* (\eta) = K_Y^* (\tau)$ whenever $\tau= \eta^{-1}$.
This is due to the relation between $I(\cdot)$ and $\Lambda^{*}(\cdot)$ described in Section~\ref{subsection:relatingLDPs}.

The last aspect of this system we wish to explore is the potential impact of channel memory and correlation among successive channel uses.
As before, we keep the probability of a bit erasure at twenty percent.
However, we vary the decay factor of the channel, $(1 - b_{12} - b_{21})$, from zero to one.
Once again, we assess performance using the mean first-passage time to an empty queue.
When the channel is memoryless, the optimal value for $K$ is $81$.
As correlation increases, more protection against erasures is beneficial and the optimal value of $K$ decreases moderately.
This enables the system to compensate for short sequences of erasures.
Still, as correlation strengthens, it becomes difficult to correct longer strings of erasures.
When this happens, the penalty of a smaller payoff produced by a low rate code begins to dominate.
In other words, attempting to recover every packet starts to be ineffective.
Rather, the code rate must be selected to transmit more information bits when the channel is favorable.
As $(1 - b_{12} - b_{21})$ approaches one, the optimal value of $K/N$ tends to one as well.
In the limit, the channel behaves much like a packet erasure model: send as many bits as possible when the channel is good and ask for retransmissions whenever the message is corrupted.
The data points that provide a basis for these findings are summarized in Table~\ref{table:OptimalK}.
\begin{table}[thb]
\begin{center}
\caption{Optimal number of information bits per codeword as a function of channel memory factor $1 - b_{12} - b_{21}$.} \label{table:OptimalK}
\begin{tabular}{|c|c|c|c|c|c|c|}
\hline
Channel & \multicolumn{2}{|c|}{Optimal Value} & \multicolumn{2}{|c|}{Mean First-Passage} &\multicolumn{2}{|c|}{Crossing} \\
Memory & \multicolumn{2}{|c|}{of $K$} & \multicolumn{2}{|c|}{Time $\mathrm{E} [H_0]$} & \multicolumn{2}{|c|}{$h_{0.95}$} \\
\hline
  & ARQ & HARQ & ARQ & HARQ & ARQ & HARQ \\
0 & 81 & 81 & 26.92 & 25.90 & 30 & 30 \\
0.5 & 77 & 78 & 28.87 & 27.79 & 32 & 32 \\
0.9 & 73 & 81 & 34.65 & 32.03 & 40 & 38 \\
0.95 & 77 & 96 & 36.68 & 31.75 & 44 & 38 \\
0.98 & 95 & 107 & 35.21 & 28.62 & 45 & 36 \\
\hline
\end{tabular}
\end{center}
\end{table}

\section{Conclusions}
\label{section:Conclusions}

This article presents a methodology for the analysis and the design of digital communication systems that operate over channels with memory.
The proposed approach is based on the time elapsed between the onset of the communication process and its termination.
Results also extend to the asymptotic decay rates of mean service and mean sojourn time.
Emphasis is on the selection of code rate for protection against erasures.
We provide a simple mathematical characterization of the first-passage time to an empty queue and the large deviations on the mean service and mean transmission time, along with a computationally efficient means to compare the performance of various implementation candidates.

The properties of coded systems are explored through a numerical study.
Optimal code rates appear robust to initial buffer conditions at the transmitter.
That is, the number of information bits to be sent from the source to the destination does not significantly affect the optimal operating point of the encoder.
Optimal operation is achieved with very similar $K$ values for mean first-passage times and various crossings of the cumulative distribution function.

For both mean service rate and mean sojourn time, it seems that the optimal operating point of a system in terms of code rate selection depends heavily on the needs of the underlying traffic.
In particular, delay-adverse applications may perform better with coarse quantization and low-rate codes.
On the other hand, delay tolerant applications may be able to use a higher rate on the same physical channel.
This phenomenon is closely related to the concept of effective capacity.

Lastly, the optimal code rate depends heavily on channel memory.
This suggest that, for systems with fixed block lengths, the channel parameters should be estimated and fed back to the encoder for optimal operation.
This naturally leads to adaptive strategies and possibly state-aware encoding schemes at the source.

\begin{appendices}

\section{Proof of Theorem~\ref{theorem:MarkovProperty1}}
\label{appendix:MarkovProperty1}

We begin this proof by introducing a convenient notation for abstract sequences.
Let $\{ a_s \}$ be a discrete-time sequence and assume that $r$ and $t$ are two integers with $r < t$.
We use $a_r^t$ to denote the subsequence $a_r, a_{r+1}, \ldots, a_t$.

Suppose $u_t = (i_t, q_t) \in \mathcal{C} \times \mathbb{N}_0$ for every $t \geq 0$.
Since $\{U_s\}_{s \geq 0}$ is a discrete-time stochastic process whose elements take on values in a finite set, it suffices to show that
\begin{equation*}
\Pr \left( U_{s+1} = u_{s+1} \middle| U_0^s = u_0^s \right)
= \Pr \left( U_{s+1} = u_{s+1} \middle| U_s=u_s \right)
\end{equation*}
in order to prove that this process is Markov.
In general, the probability on the left hand side can be expressed as
\begin{equation*}
\begin{split}
\Pr & \left( C_{(s+1)N+1} = i_{s+1} |  U_0^s = u_0^s \right) \times \\ 
&\Pr \left( Q_{s+1} = q_{s+1} |  U_0^s = u_0^s, C_{(s+1)N+1} = i_{s+1} \right) .
\end{split}
\end{equation*}
We know that the state of the channel at the onset of codeword $s+1$, labeled $C_{(s+1)N+1}$, is conditionally independent of the subsequence $Q_0^s$ and the channel states $C_1^{(s-1)N+1}$, given $C_{sN+1}$.
Thus, we get
\begin{equation*}
\begin{split}
\Pr & \left( C_{(s+1)N+1} = i_{s+1} | U_0^s = u_0^s \right) \\
&= \Pr \left( C_{(s+1)N+1} = i_{s+1} | C_{sN+1} = i_s \right) .
\end{split}
\end{equation*}
The length of the queue $Q_{s+1}$ at time $s+1$ is either $Q_s$ or $Q_s - 1$, depending on whether a codeword is successfully decoded at time~$s$.
For a non-empty queue, this depends solely on the generated codebook and the channel realizations during the transmission cycle of the codeword~$s$.
As such, we can write
\begin{equation*}
\begin{split}
\Pr & \left( Q_{s+1} = q_{s+1} |  U_0^s = u_0^s, C_{(s+1)N+1} = i_{s+1} \right) \\
&= \Pr \left( Q_{s+1} = q_{s+1} |  U_s = u_s, C_{(s+1)N+1} = i_{s+1} \right) .
\end{split}
\end{equation*}
Collecting these two results, we conclude that $\{ U_s \}$ possesses the Markov property.

\section{Proof of Proposition~\ref{proposition:PropertiesOfPf}}
\label{appendix:PropertiesOfPf}

Notice that the proposition is trivially true when $e > p$.
The only case of interest then corresponds to $e \leq p$.
We observe that, through a change in indexing, we can write
\begin{equation*}
\prod_{l=0}^{n+e-1} \left(1-2^{l-p-n}\right)
= \prod_{l=-n}^{e-1} \left( 1 - 2^{l-p} \right) .
\end{equation*}
As such, we readily see that $P_{\mathrm{f}} (p+n,e+n)$ is monotonically increasing in $n$.
The difference between this function and $P_{\mathrm{f}} (p, e)$ is obtained as follows,
\begin{equation*}
\begin{split}
P_{\mathrm{f}} & (p+n, e+n) - P_{\mathrm{f}} (p, e) \\
&= \prod_{l=0}^{e-1} \left( 1 - 2^{l-p} \right)
- \prod_{l=0}^{n+e-1} \left( 1 - 2^{l-n-p} \right) \\
&= \prod_{l=n}^{n+e-1} \left( 1 - 2^{l-n-p} \right)
- \prod_{l=0}^{n+e-1} \left( 1 - 2^{l-n-p} \right) \\
&= \prod_{l=n}^{n+e-1} \left( 1- 2^{l-n-p} \right)
\left( 1 - \prod_{l=0}^{n-1} \left( 1 - 2^{l-n-p} \right) \right) \\
&\leq 1 - \prod_{l=0}^{n-1} \left( 1 - 2^{l-n-p} \right)
\overset{(a)}{\leq} \sum_{l=0}^{n-1} 2^{l-n-p} \\
&=\sum_{l=0}^{n-1} 2^{-l-1-p}
\leq \sum_{l=0}^{\infty} 2^{-l-1-p} = 2^{-p}.
\end{split}
\end{equation*}
Step $(a)$ follows from an $n$-variable version of the inequality $1-(1-p_1)(1-p_2) \leq p_1+p_2$ where $0 \leq p_1, p_2 \leq 1$.
This concludes the demonstration.

\section{Proof of Lemma~\ref{lemma:RadiusOfConvergenceOfGT}}
\label{appendix:RadiusOfConvergenceOfGT}

When $\lambda < - \log \varrho(\mathbf{K})$, the spectral radius of the matrix $\mathbf{K} e^{\lambda}$ is strictly less than one and, consequently, the matrix $\mathbf{I} - \mathbf{K} e^{\lambda}$ is invertible.
The finiteness of $\mathbf{G}_T\left(e^{\lambda}\right)$ immediately follows.
We then turn to the alternate case, which we prove by contradiction.

Assume that, for some $\lambda \geq - \log \varrho(\mathbf{K})$, matrix $\mathbf{G}_T\left(e^{\lambda}\right)$ exists over the non-negative real numbers.
Note that this condition implies $\varrho(\mathbf{K}) > 0$.
For convenience, we wish to work with the irreducible normal form of $\mathbf{K}$~\cite{Horn0521386322}.
That is, there exists a permutation matrix $\mathbf{P}$ such that
\begin{equation*}
\tilde{\mathbf{K}} = \mathbf{P}^{\mathrm{T}} \mathbf{K} \mathbf{P}
= \begin{bmatrix}
\mathbf{\Psi}_{1} & \mathbf{\Phi}_{12} & \cdots & \mathbf{\Phi}_{1h} \\
\mathbf{0} & \mathbf{\Psi}_{2} & \cdots  & \mathbf{\Phi}_{2h} \\
\vdots & \vdots & \ddots & \vdots \\
\mathbf{0} & \mathbf{0} & \cdots & \mathbf{\Psi}_h
\end{bmatrix}
\end{equation*}
in which each $\mathbf{\Psi}_i$ is either irreducible or a zero matrix.
Of course, this reordering also affects $\mathbf{M}$,
\begin{equation*}
\tilde{\mathbf{M}} = \mathbf{P}^{\mathrm{T}} \mathbf{M} \mathbf{P} .
\end{equation*}
However, this transformation does not alter the spectrum of $\mathbf{K}$ or $\mathbf{M}$.
We note that all the states corresponding to an irreducible $\mathbf{\Psi}_i$ belong to a same communicating class, which we denote by $\mathcal{C}_i$.
Looking at the block triangular structure of $\tilde{\mathbf{K}}$, we gather that the eigenvalues of $\tilde{\mathbf{K}}$ correspond to the union of the eigenvalues of $\mathbf{\Psi}_1, \ldots, \mathbf{\Psi}_h$.
Thus, there exists an integer $j$ such that $\varrho(\mathbf{\Psi}_j) = \varrho(\mathbf{K})$.

Since matrix $\mathbf{\Psi}_j$ is non-negative and irreducible, the Perron-Frobenius theorem applies and there exists an eigenvector $\mathbf{v}$, with positive components, such that
\begin{equation*}
\mathbf{v} \mathbf{\Psi}_j = \varrho(\mathbf{\Psi}_j) \mathbf{v}
= \varrho(\mathbf{K}) \mathbf{v} .
\end{equation*}
Without loss of generality, we can assume that $\mathbf{v}$ is normalized to one.
Let $\mathbf{w}$ be a probability distribution with weight $\mathbf{v}$ over the states associated with $\mathbf{\Psi}_j$ and zero elsewhere, i.e.,
\begin{equation*}
\mathbf{w}
= \begin{bmatrix} \mathbf{0} & \cdots & \mathbf{0} & \mathbf{v}
& \mathbf{0} & \cdots & \mathbf{0} \end{bmatrix} .
\end{equation*}
Because $\mathbf{v}$ is an eigenvector of $\mathbf{\Psi}_j$, we have
\begin{equation*}
\mathbf{w} \left( \tilde{\mathbf{K}} e^{\lambda} \right)^t
= \begin{bmatrix} \mathbf{0} & \cdots & \mathbf{0}
& \left( \varrho(\mathbf{K}) e^{\lambda} \right)^t \mathbf{v}
& \mathbf{*} & \cdots & \mathbf{*} \end{bmatrix}
\end{equation*}
and, correspondingly,
\begin{equation*}
\begin{split}
\mathbf{w} & \sum_{t=0}^{\infty} \tilde{\mathbf{K}}^t e^{t \lambda}
= \sum_{t=0}^{\infty} \mathbf{w} \tilde{\mathbf{K}}^t e^{t \lambda} \\
&= \begin{bmatrix} \mathbf{0} & \cdots & \mathbf{0}
& \sum_{t = 0}^{\infty} \left( \varrho(\mathbf{K})
e^{\lambda} \right)^t \mathbf{v}
& \mathbf{*} & \cdots & \mathbf{*} \end{bmatrix} .
\end{split}
\end{equation*}
We note that the multiplicative factor $\sum_{t=0}^{\infty} \left( \varrho(\mathbf{K}) e^{\lambda} \right)^t$ is a divergent sum that increases to infinity.
In fact, all the components of $\mathbf{w} \sum_{t=0}^{\infty} \tilde{\mathbf{K}}^t e^{t \lambda}$ corresponding to states that are accessible from $\mathcal{C}_j$ must also diverge~\cite{Horn0521386322}.
Since by assumption the elements of
\begin{equation*}
\tilde{\mathbf{G}}_T \left( e^{\lambda} \right)
= \left( \sum_{t=0}^{\infty} \tilde{\mathbf{K}}^t e^{t \lambda} \right) \tilde{\mathbf{M}} e^{\lambda}
\end{equation*}
remain finite, we conclude that any state accessible from $\mathcal{C}_j$ must lie in the nullspace of $\tilde{\mathbf{M}}$.
This necessarily means that $\mathbf{w} \tilde{\mathbf{G}}_T \left( e^{\lambda} \right) = \mathbf{0}$ and, consequently, $\mathbf{w} \tilde{\mathbf{G}}_T(1) = \mathbf{0}$ because $\tilde{\mathbf{K}}$ and $\tilde{\mathbf{M}}$ are non-negative matrices.
In other words, we have created a valid probability distribution $\mathbf{w}$ for which $\mathbf{w} \tilde{\mathbf{G}}_T(1) = \mathbf{0}$.
Equivalently, in the original domain, we can rewrite this equation as $\mathbf{w} \mathbf{P}^{\mathrm{T}} \mathbf{G}_T (1) = \mathbf{0}$.
But this equation violates our assumption that $T$ is finite almost surely.
We then conclude, by contradiction, that not all entries of $\mathbf{G}_T \left(e^{\lambda}\right)$ are finite when $\lambda \geq - \log \varrho(\mathbf{K})$.

\section{Proof of Corollary~\ref{corollary:SpectralRadiusOfK}}
\label{appendix:SpectralRadiusOfK}

As a straightforward application of Lemma~\ref{lemma:RadiusOfConvergenceOfGT}, we can show that $\varrho(\mathbf{K}) < 1$.
By design, we know that $T$ is finite almost surely.
Then, from the definition of the matrix generating function $\mathbf{G}_T (z)$ in \eqref{equation:GeneratingMatrix}, we gather that
\begin{equation*}
\begin{split}
\left[ \mathbf{G}_T (1) \right]_{ij}
&= \mathrm{E} \left[ \mathbf{1}_{ \{ C_{NT+1} = j \} } | C_1 = i \right] \\
&= \Pr \left( C_{NT+1} = j | C_1 = i \right) .
\end{split}
\end{equation*}
That is, $\mathbf{G}_T (1)$ is a right stochastic matrix.

Since $\mathbf{K}$ is a substochastic matrix, we already have the relation $\varrho(\mathbf{K}) \leq 1$.
We wish to show that, in the current framework, this inequality is strict.
Suppose that $\varrho(\mathbf{K}) = 1$.
Lemma~\ref{lemma:RadiusOfConvergenceOfGT} states that, if $\lambda = - \log \varrho(\mathbf{K}) = 0$, then not all entries of $\mathbf{G}_T \left(e^0\right) \!= \mathbf{G}_T (1)$ can be finite.
In particular, $\mathbf{G}_T (1)$ cannot be a right stochastic matrix.
This leads to an obvious contradiction, which indicates that $\varrho(\mathbf{K}) < 1$, as desired.

\section{Proof of Proposition~\ref{proposition:AsymptoticMGF}}
\label{appendix:AsymptoticMGF}

For the first part of this proof, we assume that $\lambda < - \log \varrho (\mathbf{K})$.
The spectral radius of $\mathbf{K} e^{\lambda}$ is then strictly less than one and, as such, $\left( \mathbf{I} - \mathbf{K} e^{\lambda} \right)$ is invertible.
This implies that the matrix
\begin{equation*}
\mathbf{G}_T \left( e^{\lambda} \right)
= \left( \sum_{t=0}^{\infty} \mathbf{K}^t e^{t \lambda} \right) \mathbf{M} e^{\lambda}
= \left( \mathbf{I} - \mathbf{K} e^{\lambda} \right)^{-1} \mathbf{M} e^{\lambda}
\end{equation*}
is well-defined over the real numbers.
Under Assumption~\ref{assumption:IrreducibilityOfGT}, we know that $\mathbf{G}_T(1)$ is an irreducible matrix.
This readily implies that $\mathbf{G}_T\left(e^{\lambda}\right)$ is also irreducible.
We can therefore apply the Perron-Frobenius theorem~\cite[Th.~3.1.1]{Dembo0387984062}, whose asymptotic properties lead directly to $\Lambda (\lambda)$.

For the second case, we suppose that $\lambda \geq - \log \varrho (\mathbf{K})$.
By Lemma~$\ref{lemma:RadiusOfConvergenceOfGT}$, we know that at least one entry of $\mathbf{G}_T\left(e^{\lambda}\right)$ is equal to infinity.
We can use the irreducibility of this matrix to argue that each row in $\left(\mathbf{G}_T \left(e^{\lambda}\right)\right)^k$ has at least one entry that is infinite.
Since $\pi_0$ is a probability distribution,
\begin{equation*}
\mathrm{E} \left[ e^{\lambda (T_1 + \cdots + T_k)} \right]
= \pi_0 \left( \mathbf{G}_T \left(e^{\lambda}\right) \right)^k \mathbf{1} = \infty .
\end{equation*}
For any $m > k$, we have
\begin{equation*}
\begin{split}
\Lambda_m (m \lambda) &= \log \mathrm{E} \left[ e^{m \lambda Y_m} \right]
= \log \mathrm{E} \left[ e^{\lambda(T_1 + \cdots + T_m)} \right] \\
&\geq \log \mathrm{E} \left[ e^{\lambda \left( T_1 + \cdots + T_k \right)} \right] = \infty .
\end{split}
\end{equation*}
Consequently, whenever $\lambda \geq - \log \varrho (\mathbf{K})$, we get
\begin{equation*}
\Lambda(\lambda) = \lim_{m \rightarrow \infty}
\frac{1}{m} \Lambda_m (m \lambda) = \infty ,
\end{equation*}
as desired.

\section{Proof of Proposition~\ref{proposition:RelationILambda}}
\label{appendix:RelationILambda}

For the sake of completeness, we offer a brief proof for Proposition~\ref{proposition:RelationILambda}.
As an initial step for this demonstration, we establish a few key properties.
The processes $\{ Y_m \}$ and $\{ Z_s \}$ converge almost surely, i.e.,
\begin{align*}
  Y_m = \frac{1}{m} \sum_{q=1}^m T_q \xrightarrow{a.s.} \bar{T} \\
  Z_s = \frac{1}{s} \sum_{t=1}^s D_t \xrightarrow{a.s.} \bar{D},
\end{align*}
where $\bar{T}$ are $\bar{D}$ are constants.
Moreover, $\bar{T}$ and $\bar{D}$ have a reciprocal relation, i.e., $\bar{T} = 1/\bar{D}$.

Recall that process $\left\{ V_s = \left( C_{(s+1)N + 1}, D_s \right) \right\}$ is a finite-state Markov chain with irreducible transition probability matrix $\mathbf{\Pi}$.
Also, $D_s = f(V_s)$ is a (trivial) bounded function.
Then, by the ergodic theorem for Markov chains~\cite{Norris0521633966}, we have
\begin{equation*}
\Pr \left( \lim_{s \rightarrow \infty} \frac{1}{s}
\sum_{t=1}^{s} D_t = \bar{D} \right) = 1.
\end{equation*}
Let $\Omega_1$ be the subset of $\Omega$ defined by
\begin{equation*}
\Omega_1 = \left\{ \omega : \frac{1}{s}
\sum_{t=1}^{s} D_t (\omega) \rightarrow \bar{D} \right\} .
\end{equation*}
Clearly, for any $\omega \in \Omega_1$, we necessarily have
\begin{equation*}
N(s, \omega) = \sum_{t=1}^{s} D_t (\omega) \rightarrow \infty .
\end{equation*}

Consider the empirical average defined by
\begin{equation}
\label{equation:EmpiricalAverageT}
\frac{1}{m} \sum_{q=1}^{m} T_q .
\end{equation}
We wish to show that this sequence converges almost surely to $1/\bar{D}$ as $m$ increases to infinity.
For any $\omega \in \Omega_1$, we have
\begin{equation*}
\sum_{q=1}^{N(s, \omega)} T_q (\omega)
\leq s \leq
\sum_{q=1}^{N(s, \omega)+1} T_q (\omega) .
\end{equation*}
As such, we get the inequality
\begin{equation*}
\frac{1}{N(s, \omega)} \sum_{q=1}^{N(s, \omega)} T_q (\omega)
\leq \frac{s}{N(s, \omega)} \rightarrow \frac{1}{\bar{D}} .
\end{equation*}
In a similar fashion, we obtain
\begin{equation*}
\begin{split}
\frac{1}{N(s, \omega)+1} & \sum_{q=1}^{N(s, \omega)+1} T_q (\omega)
\geq \frac{s}{N(s, \omega)+1} \\
&= \frac{N(s, \omega)}{N(s, \omega)+1} \frac{s}{N(s, \omega)}
\rightarrow \frac{1}{\bar{D}} .
\end{split}
\end{equation*}
It follows that, for any $\omega \in \Omega_1$, we get
\begin{equation}
\label{equation:ExtendedEmpiricalAverageT}
\frac{1}{N(s, \omega)} \sum_{q=1}^{N(s, \omega)} T_q (\omega)
\rightarrow \frac{1}{\bar{D}} .
\end{equation}

To complete the proof, we must connect this result to our original sequence~\eqref{equation:EmpiricalAverageT}.
We emphasize that, for any $\omega \in \Omega_1$ and for any $m \in \mathbb{N}$, there exists $s$ such that $N(s, \omega) = m$ because $N(s, \omega)$ increases by at most one at every step.
It follows that \eqref{equation:EmpiricalAverageT} is a subsequence of convergent sequence \eqref{equation:ExtendedEmpiricalAverageT}.
They must then share the same limit.
Collecting these results, we gather that
\begin{equation*}
\Pr \left( \lim_{m \rightarrow \infty} \frac{1}{m} \sum_{q=1}^{m} T_q = \frac{1}{\bar{D}} \right) = 1.
\end{equation*}
As a side note, it is possible to show that
\begin{align*}
\bar{D} &= \mathrm{E}_{\pi_D} \left[ D_t \right]
= \pi_D \mathbf{M} \mathbf{1} \\
\bar{T} & =\mathrm{E}_{\pi_T} \left[ T_q \right]
= \pi_T \left[ \lim_{\lambda \uparrow 0} \frac{d}{d \lambda}
\mathbf{G}_T \left( e^{\lambda} \right) \right] \mathbf{1} ,
\end{align*}
where $\frac{d}{d \lambda} \mathbf{G}_T(e^{\lambda})$ denotes the entrywise derivative.
Above, $\pi_D$ and $\pi_T$ represent the invariant distributions of the channel and the stochastic matrix $\mathbf{G}_T(1)$, respectively.

Our strategy to finish this proof is to establish the claimed result for rational numbers, and then invoke continuity to get a full characterization.
From our hypotheses, we know that the rate functions $\Lambda^*(\cdot)$ and $I(\cdot)$ are finite in the open intervals $(1,\infty)$ and $(0,1)$, respectively.
We note that these functions are also convex over these intervals and, hence, continuous.
Let $r = p/q$, where $p, q \in \mathbb{N}$, be a rational number less than one.
Recall that $I(\cdot)$ is convex and, therefore, continuous over $(0,1)$.
Then, for every $\epsilon > 0$, there exists $\delta > 0$ such that
\begin{equation*}
\begin{split}
&- I (r) - \epsilon
\leq \liminf_{n \rightarrow \infty} \frac{1}{np} \log 
\Pr \left( Z_{np} \in ( r - \delta, r + \delta ) \right) \\
&\leq \limsup_{n \rightarrow \infty} \frac{1}{np} \log 
\Pr \left( Z_{np} \in ( r - \delta, r + \delta ) \right)
\leq - I (r) + \epsilon .
\end{split}
\end{equation*}
Taking the limit as $\delta \rightarrow 0$, we get
\begin{equation*}
\begin{split}
&\lim_{\delta \rightarrow 0} \liminf_{n \rightarrow \infty} \frac{1}{np} \log 
\Pr \left( Z_{np} \in ( r - \delta, r + \delta ) \right) \\
&= \lim_{\delta \rightarrow 0} \limsup_{n \rightarrow \infty} \frac{1}{np} \log 
\Pr \left( Z_{np} \in ( r - \delta, r + \delta ) \right)
= - I(r) .
\end{split}
\end{equation*}
A similar argument applies to $\{ Y_m \}$.
Noting that $q / p \in (1, \infty)$, we gather that $\Lambda^* (\cdot)$ is continuous in a neighborhood of $1/r$.
Then, for every $\epsilon > 0$, there exists $\delta > 0$ such that
\begin{equation*}
\begin{split}
- \Lambda^* & \left( \frac{1}{r} \right) - \epsilon \\
&\leq \liminf_{n \rightarrow \infty} \frac{1}{nq} \log 
\Pr \left( Y_{nq} \in \left( \frac{1}{r} - \delta, \frac{1}{r} + \delta \right) \right) \\
&\leq \limsup_{n \rightarrow \infty} \frac{1}{nq} \log 
\Pr \left( Y_{nq} \in \left( \frac{1}{r} - \delta, \frac{1}{r} + \delta \right) \right) \\
&\leq - \Lambda^* \left( \frac{1}{r} \right) + \epsilon .
\end{split}
\end{equation*}
As before, this implies that
\begin{equation*}
\begin{split}
&\lim_{\delta \rightarrow 0} \liminf_{n \rightarrow \infty} \frac{1}{nq} \log 
\Pr \left( Y_{nq} \in \left( \frac{1}{r} - \delta, \frac{1}{r} + \delta \right) \right) \\
&= \lim_{\delta \rightarrow 0} \limsup_{n \rightarrow \infty} \frac{1}{nq} \log 
\Pr \left( Y_{nq} \in \left( \frac{1}{r} - \delta, \frac{1}{r} + \delta \right) \right) \\
&= - \Lambda^* \left( \frac{1}{r} \right) .
\end{split}
\end{equation*}
We stress that the rate functions $\Lambda^*(\cdot)$ and $I(\cdot)$ vanish at $\bar{T}$ and $\bar{D}$, respectively.

At this point, we need to consider two separate cases.
First, suppose $r < \bar{D}$.
We know that $I(\cdot)$ is a non-increasing function over interval $\left[ 0, \bar{D} \right)$ (see, e.g., \cite[Lemma 2.2.5]{Dembo0387984062}).
Also, in an analogous manner, rate function $\Lambda^* (\cdot)$ is non-decreasing over $\left( \bar{T}, \infty \right)$.
Leveraging \eqref{equation:EquivalentEvents}, we can write
\begin{equation*}
\Pr \left( \frac{T_1 + \cdots + T_{pn}}{pn} > \frac{q}{p} \right)
= \Pr \left( \frac{D_1 + \cdots + D_{qn}}{qn} < \frac{p}{q}\right) .
\end{equation*}
By letting $n$ go to infinity, we obtain
\begin{equation*}
\inf_{x \in \left[ \frac{1}{r}, \infty \right)} r \Lambda^* (x)
= \inf_{x \in (0, r] } I(x) .
\end{equation*}
Using the monotonic properties of these rate functions over the prescribed intervals, we get
\begin{equation*}
r \Lambda^* \left( \frac{1}{r} \right)
= \inf_{x \in \left[ \frac{1}{r}, \infty \right)} r \Lambda^* (x)
= \inf_{x \in (0, r] } I(x)
= I(r) ,
\end{equation*}
as desired.

For the second case, assume $r > \bar{D}$.
Under this constraint, the monotonic properties of the rate functions are reversed.
That is, $I(\cdot)$ is non-decreasing over $\left( \bar{D}, 1 \right)$ and $\Lambda^* (\cdot)$ is non-increasing over $\left( 0, \bar{T} \right)$.
Using these relations and the set equalities
\begin{equation*}
\Pr \left( \frac{T_1 + \cdots + T_{pn}}{pn} < \frac{q}{p} \right)
= \Pr \left( \frac{D_1 + \cdots + D_{qn}}{qn} > \frac{p}{q}\right) ,
\end{equation*}
we can write
\begin{equation*}
r \Lambda^* \left( \frac{1}{r} \right)
= \inf_{x \in \left( 0, \frac{1}{r} \right]} r \Lambda^* (x)
= \inf_{x \in [r, \infty) } I(x)
= I(r) .
\end{equation*}

Collecting these results, we deduce that $I(x) = x \Lambda^* \left( \frac{1}{x} \right)$ whenever $x \in \mathbb{Q} \cap (0,1)$.
Since the rational numbers are dense in $(0,1)$ and the two rate functions are continuous, this equality must also hold for any real number in $(0,1)$.

\end{appendices}

\nocite{draper2009rateless}
\bibliographystyle{IEEEtran}
\bibliography{firstpassagetime1}

\end{document}

%% file: Figures/erasurechannel.tex
\begin{tikzpicture}
[node distance = 12mm, draw=black, thick, >=stealth',
  state/.style={circle, inner sep = 0pt, minimum size = 8mm, draw=black}]

\node[state] (c2) at (0,2) {\footnotesize $2$};
\draw [->] (c2) to [out=125, in=175, looseness=5] (c2);

\node[state] (c1) at (0,0) {\footnotesize $1$}
    edge[<-, bend right=45] node[auto,swap]{$b_{21}$} (c2)
    edge[->, bend left=45] node[auto]{$b_{12}$}(c2);
\draw [->] (c1) to [out=-55, in=-5, looseness=5] (c1);

\node[coordinate] (c2t1) at (2,2) [yshift=12, label=left:$1$]{};
\node[coordinate] (c2r1) [right=of c2t1, label=right:$1$]{}
    edge[<-] (c2t1);
\node[coordinate] (c2t0) at (2,2) [yshift=-12, label=left:$0$]{};
\node[coordinate] (c2r0)[right=of c2t0, label=right:$0$]{}
    edge[<-](c2t0);
\node[coordinate] (c2te) at (2,2) [label=right:$\varepsilon_2$]{};
\node[coordinate] (c2re)[right=of c2te, label=right:$e$]{}
    edge[<-] (c2t1)
    edge[<-] (c2t0);

\node[coordinate] (c1t1) at (2,0) [yshift=12, label=left:$1$]{};
\node[coordinate] (c1r1)[right=of c1t1, label=right:$1$]{}
    edge[<-](c1t1);
\node[coordinate] (c1t0) at (2,0) [yshift=-12, label=left:$0$]{};
\node[coordinate] (c1r0)[right=of c1t0, label=right:$0$]{}
    edge[<-](c1t0);
\node[coordinate] (c1te) at (2,0) [label=right:$\varepsilon_1$]{};
\node[coordinate] (c1re)[right=of c1te, label=right:$e$]{}
    edge[<-] (c1t1)
    edge[<-] (c1t0);
\end{tikzpicture}

%% file: Figures/packeterasurequeue.tex
\begin{tikzpicture}
[node distance = 12mm, draw=black, thick, >=stealth',
  state/.style={circle, inner sep = 0pt, minimum size = 8mm, draw=black},
  hiddenstate/.style={circle, minimum size = 8mm}]
  
  \node[state] (c20) at (0,2) {\footnotesize $2,0$};
  \node[state] (c10) at (0,0) {\footnotesize $1,0$};
  \node[state] (c21) at (2,2) {\footnotesize $2,1$};
  \node[state] (c11) at (2,0) {\footnotesize $1,1$};
  \node[hiddenstate] (c2i) at (4,2) {};
  \node[hiddenstate] (c1i) at (4,0) {};
  \node[state] (c2m) at (6,2) {\footnotesize $2,m$};
  \node[state] (c1m) at (6,0) {\footnotesize $1,m$};

  \draw[->, looseness=5] (c2m) to [in=25,out=65] (c2m);
  \draw[->, looseness=5] (c1m) to [in=-25,out=-65] (c1m);
  \draw[->, looseness=5] (c20) to [in=25,out=65] (c20);
  \draw[->, looseness=5] (c10) to [in=-25,out=-65] (c10);
  \draw[->, looseness=5] (c21) to [in=25,out=65] (c21);
  \draw[->, looseness=5] (c11) to [in=-25,out=-65] (c11);

  \draw[->] (c2m) to [bend left=25] (c1m);
  \draw[->] (c1m) to [bend left=25] (c2m);
  \draw[->] (c21) to [bend left=25] (c11);
  \draw[->] (c11) to [bend left=25] (c21);
  \draw[->] (c20) to [bend left=25] (c10);
  \draw[->] (c10) to [bend left=25] (c20);
  
  \draw[loosely dashed] (c2m) to (c2i);
  \draw[loosely dashed] (c2m) to (c1i);
  \draw[loosely dashed] (c1m) to (c2i);
  \draw[loosely dashed] (c1m) to (c1i);

  \draw[->, loosely dashed] (c2i) to (c21);
  \draw[->, loosely dashed] (c2i) to (c11);
  \draw[->, loosely dashed] (c1i) to (c21);
  \draw[->, loosely dashed] (c1i) to (c11);

  \draw[->] (c21) to (c20);
  \draw[->] (c21) to (c10);
  \draw[->] (c11) to (c20);
  \draw[->] (c11) to (c10);

  \node (c2i) at (4,1) {\Huge $\dots$};
\end{tikzpicture}

%% file: Figures/packeterasureblock.tex
\begin{tikzpicture}
[node distance = 12mm, draw=black, thick, >=stealth',
  state/.style={circle, inner sep = 0pt, minimum size = 8mm, draw=black}]
  
  \node[state] (c20) at (0,2) {\footnotesize $2,0$};
  \node[state] (c10) at (0,0) {\footnotesize $1,0$};
  \node[state] (c21) at (2,2) {\footnotesize $2,1$};
  \node[state] (c11) at (2,0) {\footnotesize $1,1$};

  \draw[->, looseness=5] (c20) to [in=25,out=65] node[above]{\footnotesize 1} (c20);
  \draw[->, looseness=5] (c10) to [in=-25,out=-65] node[below]{\footnotesize 1} (c10);
  \draw[->, looseness=5] (c21) to [in=25,out=65] (c21);
  \draw[->, looseness=5] (c11) to [in=-25,out=-65] (c11);

  \draw[->] (c21) to [bend left=25] (c11);
  \draw[->] (c11) to [bend left=25] (c21);
  \draw[->] (c21) to (c20);
  \draw[->] (c21) to (c10);
  \draw[->] (c11) to (c20);
  \draw[->] (c11) to (c10);
\end{tikzpicture}

%% file: Figures/MeanHittingTime.tex
%
%
\begin{tikzpicture}

\begin{axis}[%
scale only axis,
width=\tikzwidth,
height=\tikzheight,
xmin=50, xmax=90,
ymin=31, ymax=42,
xlabel={Information Bits (K)},
ylabel={Mean Hitting Time},
xmajorgrids,
ymajorgrids,
zmajorgrids,
legend entries={\footnotesize{Hybrid ARQ-Upper bound},\footnotesize{Hybrid ARQ-Lower bound},\footnotesize{ARQ}},
legend style={nodes=right}]
\addplot [
color=black,
solid,
line width=1.5pt
]
coordinates{
 (50,41.689)(51,40.9896)(52,40.325)(53,39.6938)(54,39.095)(55,38.5268)(56,37.9886)(57,37.479)(58,36.9971)(59,36.5416)(60,36.1122)(61,35.7082)(62,35.3283)(63,34.9718)(64,34.6389)(65,34.3284)(66,34.0398)(67,33.7725)(68,33.5263)(69,33.3003)(70,33.0946)(71,32.9081)(72,32.7409)(73,32.5923)(74,32.4621)(75,32.3495)(76,32.2546)(77,32.1766)(78,32.1154)(79,32.0702)(80,32.0404)(81,32.0265)(82,32.0267)(83,32.0416)(84,32.07)(85,32.1115)(86,32.1658)(87,32.232)(88,32.3091)(89,32.3975)(90,32.4955) 
};

\addplot [
color=black,
only marks,
mark=o,
mark size=1.5pt,
mark options={solid}
]
coordinates{
 (50,41.689)(51,40.9896)(52,40.325)(53,39.6938)(54,39.095)(55,38.5268)(56,37.9886)(57,37.479)(58,36.9972)(59,36.5417)(60,36.1122)(61,35.7082)(62,35.3283)(63,34.9718)(64,34.6389)(65,34.3284)(66,34.0398)(67,33.7726)(68,33.5263)(69,33.3003)(70,33.0946)(71,32.9081)(72,32.7409)(73,32.5923)(74,32.4622)(75,32.3496)(76,32.2546)(77,32.1766)(78,32.1154)(79,32.0702)(80,32.0405)(81,32.0265)(82,32.0267)(83,32.0416)(84,32.07)(85,32.1116)(86,32.1659)(87,32.2321)(88,32.3092)(89,32.3976)(90,32.4957) 
};

\addplot [
color=blue,
dashed,
line width=1.5pt
]
coordinates{
 (50,41.7339)(51,41.0419)(52,40.3859)(53,39.7646)(54,39.1772)(55,38.6219)(56,38.0986)(57,37.6061)(58,37.1436)(59,36.7103)(60,36.3061)(61,35.9308)(62,35.5835)(63,35.264)(64,34.9729)(65,34.7098)(66,34.4748)(67,34.2681)(68,34.0901)(69,33.9409)(70,33.8217)(71,33.7326)(72,33.6747)(73,33.649)(74,33.6566)(75,33.6986)(76,33.7768)(77,33.8928)(78,34.0489)(79,34.2468)(80,34.489)(81,34.7793)(82,35.1197)(83,35.5152)(84,35.9692)(85,36.4868)(86,37.074)(87,37.7365)(88,38.4816)(89,39.3189)(90,40.257) 
};

\end{axis}
\end{tikzpicture}

%% file: Figures/VarHittingTime.tex
%
%
\begin{tikzpicture}

\begin{axis}[%
scale only axis,
width=\tikzwidth,
height=\tikzheight,
xmin=50, xmax=90,
ymin=5, ymax=40,
xlabel={Information Bits (K)},
ylabel={Variance of Hitting Time},
xmajorgrids,
ymajorgrids,
zmajorgrids,
legend entries={\footnotesize{Hybrid ARQ},\footnotesize{ARQ}},
legend style={nodes=right}]

\addplot [
color=black,
line width=1.5pt,
solid
]
coordinates{
 (50,5.50151)(51,5.42811)(52,5.37459)(53,5.3445)(54,5.31476)(55,5.31157)(56,5.3057)(57,5.31489)(58,5.32825)(59,5.36729)(60,5.40601)(61,5.45145)(62,5.50828)(63,5.58437)(64,5.6593)(65,5.73779)(66,5.82718)(67,5.92432)(68,6.02664)(69,6.1431)(70,6.25199)(71,6.37595)(72,6.50127)(73,6.62866)(74,6.75771)(75,6.89617)(76,7.03213)(77,7.16559)(78,7.29533)(79,7.42859)(80,7.56764)(81,7.68297)(82,7.81249)(83,7.92475)(84,8.03702)(85,8.13723)(86,8.2251)(87,8.30632)(88,8.38531)(89,8.43432)(90,8.4765) 
};

\addplot [
color=blue,
line width=1.5pt,
dashed
]
coordinates{
 (50,5.64633)(51,5.59685)(52,5.57089)(53,5.57255)(54,5.57924)(55,5.6179)(56,5.6599)(57,5.72386)(58,5.79975)(59,5.91026)(60,6.03034)(61,6.16836)(62,6.33047)(63,6.5263)(64,6.73695)(65,6.96919)(66,7.23279)(67,7.52708)(68,7.8523)(69,8.22093)(70,8.61409)(71,9.05938)(72,9.5471)(73,10.0831)(74,10.6729)(75,11.3314)(76,12.0534)(77,12.8473)(78,13.7215)(79,14.6952)(80,15.7843)(81,16.9694)(82,18.3117)(83,19.7942)(84,21.4608)(85,23.3236)(86,25.4143)(87,27.7779)(88,30.4661)(89,33.4945)(90,36.9606) 
};

\end{axis}
\end{tikzpicture}

%% file: Figures/CrossingsCDF.tex
%
%
\begin{tikzpicture}

\begin{axis}[%
scale only axis,
width=\tikzwidth,
height=\tikzheight,
xmin=50, xmax=90,
ymin=32, ymax=52,
xlabel={Information Bits (K)},
ylabel={Transmission Attempts},
xmajorgrids,
ymajorgrids,
zmajorgrids,
legend entries={\footnotesize{ARQ},\footnotesize{Hybrid ARQ}},
legend style={at={(0.03,0.97)},anchor=north west,nodes=right}]

\addplot [
color=blue,
line width=1.5 pt,
dashed
]
coordinates{
 (50,42)(51,42)(52,41)(53,40)(54,40)(55,39)(56,39)(57,38)(58,38)(59,37)(60,37)(61,36)(62,36)(63,36)(64,36)(65,35)(66,35)(67,35)(68,35)(69,34)(70,34)(71,34)(72,34)(73,34)(74,34)(75,34)(76,34)(77,34)(78,34)(79,34)(80,35)(81,35)(82,35)(83,36)(84,36)(85,37)(86,37)(87,38)(88,38)(89,39)(90,40) 
};

\addplot [
color=black,
line width=1.5 pt,
solid
]
coordinates{
 (50,42)(51,42)(52,41)(53,40)(54,40)(55,39)(56,39)(57,38)(58,38)(59,37)(60,37)(61,36)(62,36)(63,36)(64,35)(65,35)(66,35)(67,34)(68,34)(69,34)(70,34)(71,33)(72,33)(73,33)(74,33)(75,33)(76,33)(77,33)(78,33)(79,33)(80,33)(81,33)(82,33)(83,33)(84,33)(85,33)(86,33)(87,33)(88,33)(89,33)(90,33) 
};

\addplot [
color=blue,
line width=1.5 pt,
dashed
]
coordinates{
 (50,47)(51,46)(52,45)(53,45)(54,44)(55,44)(56,43)(57,43)(58,42)(59,42)(60,42)(61,41)(62,41)(63,41)(64,40)(65,40)(66,40)(67,40)(68,40)(69,40)(70,40)(71,40)(72,40)(73,40)(74,40)(75,41)(76,41)(77,41)(78,42)(79,42)(80,42)(81,43)(82,44)(83,44)(84,45)(85,46)(86,47)(87,48)(88,49)(89,50)(90,52) 
};

\addplot [
color=black,
line width=1.5 pt,
solid
]
coordinates{
 (50,47)(51,46)(52,45)(53,45)(54,44)(55,43)(56,43)(57,42)(58,42)(59,41)(60,41)(61,41)(62,40)(63,40)(64,40)(65,39)(66,39)(67,39)(68,39)(69,39)(70,38)(71,38)(72,38)(73,38)(74,38)(75,38)(76,38)(77,38)(78,38)(79,38)(80,38)(81,38)(82,38)(83,38)(84,38)(85,38)(86,38)(87,38)(88,38)(89,38)(90,38) 
};

\end{axis}
\end{tikzpicture}

%% file: Figures/RateFunctionService.tex
%
%
\begin{tikzpicture}

\definecolor{mycolor1}{rgb}{0,0.5,0}

\begin{axis}[%
scale only axis,
width=\tikzwidth,
height=\tikzheight,
xmin=20, xmax=100,
ymin=0, ymax=0.012,
xlabel={$\text{Information Bits (K)}$},
ylabel={$\frac{1}{N}I\left(\frac{N}{K}\eta\right)$},
y tick label style={/pgf/number format/.cd,fixed,fixed zerofill,precision=3},
xmajorgrids,
ymajorgrids,
legend entries={$\eta=0.200$,$\eta=0.275$,$\eta=0.350$,$\eta=0.425$},
legend style={nodes=right}]
\addplot [
color=blue,
solid,
line width=1.0pt,
mark=o,
mark options={solid}
]
coordinates{ (20.5,2.22045e-16) (22,2.22045e-16) (26,0.00322524) (30,0.00670014) (34,0.00888623) (38,0.0101) (42,0.0106166) (46,0.0106354) (50,0.0102994) (54,0.00971233) (58,0.0089511) (62,0.00807417) (66,0.00712704) (70,0.00614607) (74,0.00516124) (78,0.00419809) (82,0.00327932) (86,0.00242611) (90,0.00165948) (94,0.00100189) (98,0.000479752) (99.5,0.000347862)
};

\addplot [
color=mycolor1,
solid,
line width=1.0pt,
mark=square,
mark options={solid}
]
coordinates{ (20.5,2.22045e-16) (22,2.22045e-16) (26,2.22045e-16) (30,2.22045e-16) (34,0.00112724) (38,0.003099) (42,0.00446729) (46,0.00527032) (50,0.00562704) (54,0.00564571) (58,0.00541427) (62,0.00500276) (66,0.00446725) (70,0.00385337) (74,0.00319909) (78,0.00253696) (82,0.00189596) (86,0.00130313) (90,0.000785277) (94,0.000371109) (98,9.45084e-05) (99.5,5.90678e-05)
};

\addplot [
color=red,
solid,
line width=1.0pt,
mark size=2.5pt,
mark=x,
mark options={solid}
]
coordinates{ (20.5,2.22045e-16) (22,2.22045e-16) (26,2.22045e-16) (30,2.22045e-16) (34,2.22045e-16) (38,2.22045e-16) (42,0.000246617) (46,0.00121394) (50,0.00196619) (54,0.0024108) (58,0.00258716) (62,0.00255215) (66,0.00236015) (70,0.00205921) (74,0.00169122) (78,0.00129311) (82,0.000898391) (86,0.000538706) (90,0.000245763) (94,5.3814e-05) (98,2.22045e-16) (99.5,2.22045e-16)
};

\addplot [
color=black,
solid,
line width=1.0pt,
mark=diamond,
mark options={solid}
]
coordinates{ (20.5,2.22045e-16) (22,2.22045e-16) (26,2.22045e-16) (30,2.22045e-16) (34,2.22045e-16) (38,2.22045e-16) (42,2.22045e-16) (46,2.22045e-16) (50,3.03305e-07) (54,0.00028498) (58,0.000602805) (62,0.00078644) (66,0.000832297) (70,0.000768109) (74,0.000628551) (78,0.000448834) (82,0.000263309) (86,0.000105926) (90,1.17113e-05) (94,2.22045e-16) (98,2.22045e-16) (99.5,2.22045e-16)
};

\end{axis}
\end{tikzpicture}

%% file: Figures/RateFunctionHitting.tex
%
%
\begin{tikzpicture}

\definecolor{mycolor1}{rgb}{0,0.5,0}

\begin{axis}[%
scale only axis,
width=\tikzwidth,
height=\tikzheight,
xmin=20, xmax=100,
ymin=0, ymax=0.06,
xlabel={$\text{Information Bits (K)}$},
ylabel={$\frac{1}{K}\Lambda^{*}\left(\frac{K}{N}\tau\right)$},
y tick label style={/pgf/number format/.cd,fixed,fixed zerofill,precision=3},
ytick={0,0.010,0.020,0.030,0.040,0.050,0.060},
xmajorgrids,
ymajorgrids,
legend entries={$\tau=5.000$,$\tau=3.636$,$\tau=2.857$,$\tau=2.353$},
legend style={nodes=right}]
\addplot [
color=blue,
solid,
line width=1.0pt,
mark=o,
mark options={solid}
]
coordinates{ (20.5,2.22045e-16) (22,2.22045e-16) (26,0.0161262) (30,0.0335007) (34,0.0444312) (38,0.0505002) (42,0.053083) (46,0.0531772) (50,0.0514972) (54,0.0485617) (58,0.0447555) (62,0.0403709) (66,0.0356352) (70,0.0307304) (74,0.0258062) (78,0.0209905) (82,0.0163966) (86,0.0121306) (90,0.00829739) (94,0.00500945) (98,0.00239876) (99.5,0.00173931)
};

\addplot [
color=mycolor1,
solid,
line width=1.0pt,
mark=square,
mark options={solid}
]
coordinates{ (20.5,2.22045e-16) (22,2.22045e-16) (26,2.22045e-16) (30,2.22045e-16) (34,0.00409906) (38,0.0112691) (42,0.0162447) (46,0.0191648) (50,0.020462) (54,0.0205298) (58,0.0196882) (62,0.0181918) (66,0.0162446) (70,0.0140123) (74,0.0116331) (78,0.00922532) (82,0.00689442) (86,0.00473866) (90,0.00285555) (94,0.00134949) (98,0.000343667) (99.5,0.000214792)
};

\addplot [
color=red,
solid,
line width=1.0pt,
mark size=2.5pt,
mark=x,
mark options={solid}
]
coordinates{ (20.5,2.22045e-16) (22,2.22045e-16) (26,2.22045e-16) (30,2.22045e-16) (34,2.22045e-16) (38,2.22045e-16) (42,0.000704619) (46,0.0034684) (50,0.00561769) (54,0.00688801) (58,0.00739188) (62,0.00729185) (66,0.00674328) (70,0.00588345) (74,0.00483205) (78,0.00369461) (82,0.00256683) (86,0.00153916) (90,0.00070218) (94,0.000153754) (98,2.22045e-16) (99.5,2.22045e-16)
};

\addplot [
color=black,
solid,
line width=1.0pt,
mark=diamond,
mark options={solid}
]
coordinates{ (20.5,2.22045e-16) (22,2.22045e-16) (26,2.22045e-16) (30,2.22045e-16) (34,2.22045e-16) (38,2.22045e-16) (42,2.22045e-16) (46,2.22045e-16) (50,7.13658e-07) (54,0.000670542) (58,0.00141836) (62,0.00185045) (66,0.00195835) (70,0.00180731) (74,0.00147894) (78,0.00105608) (82,0.000619551) (86,0.000249238) (90,2.75561e-05) (94,2.22045e-16) (98,2.22045e-16) (99.5,2.22045e-16)
};

\end{axis}
\end{tikzpicture}